\documentclass[1p]{elsarticle}

\usepackage{ulem}
\usepackage{graphicx,float,amsmath,amssymb,mathrsfs}
\usepackage[english,francais]{babel}
\usepackage{url,hyperref}
\usepackage[usenames,dvipsnames]{color}

  \definecolor{blue}{rgb}{0,0,1}
  \definecolor{green}{rgb}{0,.6,0}
  \definecolor{red}{rgb}{1,0,0}
  \definecolor{vio}{rgb}{1,0,1}
  \definecolor{uv}{rgb}{0.5,0,0.5}
  \definecolor{ama}{rgb}{0.3,0.3,0.3}

\definecolor{M_Beige}         {rgb}{0.96 , 0.96 , 0.86}

\definecolor{M_Brown}         {rgb}{0.65 , 0.16 , 0.16}

\definecolor{M_Gold}          {rgb}{1.00 , 0.84 , 0.00}

\definecolor{M_LemonChiffon}  {rgb}{1.00 , 0.98 , 0.80}

\definecolor{M_Orange}        {rgb}{1.00 , 0.60 , 0.00}

\definecolor{M_Pink}          {rgb}{0.80 , 0.55 , 0.60}

\definecolor{M_Violet}          {rgb}{0.83 , 0.21 , 0.93}

\definecolor{M_Green}          {rgb}{0.2 , 0.6 , 0.2}

\definecolor{M_Gray}          {rgb}{0.7 , 0.7 , 0.7}

\definecolor{M_BluPal}          {rgb}{0.7 , 0.7 , 0.9}

\renewcommand{\geq}{\geqslant}

\newcommand{\ket}[1]{|\kern.3ex#1\kern.3ex\rangle}
\newcommand{\bra}[1]{\langle\kern.3ex #1 \kern.3ex|}
\newcommand{\scalar}[2]{\langle\kern.3ex{#1}\kern.3ex|\kern.3ex{#2}\kern.3ex\rangle}

\newcommand \be  {\begin{equation}}
\newcommand \bea {\begin{eqnarray} \nonumber }
\newcommand \ee  {\end{equation}}
\newcommand \eea {\end{eqnarray}}

\newcommand{\corr}[1]{\langle#1\rangle}
\DeclareMathOperator{\Str}{Str}
\DeclareMathOperator{\diag}{diag}
\renewcommand{\Re}{\mathop{\rm Re}}
\renewcommand{\Im}{\mathop{\rm Im}}

\def\XXint#1#2#3{{\setbox0=\hbox{$#1{#2#3}{\int}$}
\vcenter{\hbox{$#2#3$}}\kern-.5\wd0}}

\begin{document}

\selectlanguage{english}

%%%%%%%%%%%%%%%%%%%%%%%%%%%%%%%%%%%%%%%%%%%%%%%%%%%%%%%%%%%%%%%%%%%%%%%%%%%%%
\renewcommand{\labelitemi}{$\bullet$}
\renewcommand{\labelitemii}{$\star$}
%%%%%%%%%%%%%%%%%%%%%%%%%%%%%%%%%%%%%%%%%%%%%%%%%%%%%%%%%%%%%%%%%%%%%%%%%%%%%

\title{Resonances in a single-lead reflection from a disordered medium:  $\sigma$-model approach}

\author{Yan V. Fyodorov}
\ead{yan.fyodorov@kcl.ac.uk}
\address{King's College London, Department of Mathematics, London  WC2R 2LS, United Kingdom}

\author{Mikhail A. Skvortsov}
\ead{skvor@itp.ac.ru}
\address{L. D. Landau Institute for Theoretical Physics, 142432 Chernogolovka, Russia}

\author{Konstantin S. Tikhonov}
\ead{tikhonov.konstantin@zoho.com}
\address{Capital Fund Management, 23 rue de l'Universit\'e, 75007 Paris, France}

\begin{abstract}
Using the framework of supersymmetric non-linear $\sigma$-model we develop a general non-perturbative characterisation of universal features of the density $\rho(\Gamma)$ of the imaginary parts (``width'') for $S$-matrix poles (``resonances'') describing waves incident and reflected from a disordered medium via  $M$-channel waveguide/lead. Explicit expressions for $\rho(\Gamma)$ are derived for several instances of systems with broken time-reversal invariance, in particular for quasi-1D and 3D media. In the case of perfectly coupled lead with a few channels ($M\sim 1$) the most salient features are tails $\rho(\Gamma)\sim \Gamma^{-1}$ for narrow resonances reflecting exponential localization and $\rho(\Gamma)\sim \Gamma^{-2}$ for broad resonances reflecting states located in the vicinity of the attached wire.
For multimode quasi 1D wires with $M\gg 1$, an intermediate asymptotics $\rho(\Gamma)\sim \Gamma^{-3/2}$ is shown to emerge reflecting diffusive nature of decay into wide enough contacts.
\end{abstract}


\maketitle

{\small
\setcounter{tocdepth}{2}
\tableofcontents
}

%%%%%%%%%%%%%%%%%%%%%%%%%%%%%%%%%%%%%%%%%%%%%%%%%%%%%%%%%%%%%%%%%%%%%%%%%%%%%%%%%%%%%%%%%%
%%%%%%%%%%%%%%%%%%%%%%%%%%%%%%%%%%%%%%%%%%%%%%%%%%%%%%%%%%%%%%%%%%%%%%%%%%%%%%%%%%%%%%%%%%

\section{Introduction}
%%%%%%%%%%%%%%%%%%%%%%%%%%%%%%%%%%%%%%%%%%%%%%%%%%%%%%%%%%%%%%%%%%%%%%%%%%%%%%%%%%%%%%%%%%%%%%%%%%%%%%%%%%%%%

\subsection{Statistics of resonances in scattering from disordered media: discussion of the problem}

Quantitative statistical characterization of various aspects of wave scattering from samples of disordered medium with randomly distributed impurities remains an active research area across  Theoretical, Experimental, and Engineering Physics, see e.g.\ Refs.\  \cite{RN99,Fyo05_rev,Wiersma2013,Tian2013_rev,Rotter_Gigan_rev}.  Many features of such scattering are common both to classical waves and a single quantum particle incident on the sample and eventually escaping from it, the processes most conveniently characterized in terms of  frequency or energy-dependent unitary scattering matrix $S(E)$. One of problems in this area is to describe statistics of resonances, defined as poles ${\cal E}_n=E_n-i\Gamma_n$ of  $S(E)$ in the lower half of the complex energy plane ${\cal E}=E-i\Gamma,\, \Gamma>0$ \cite{resbook}. In particular, the imaginary parts $\Gamma_n$ (traditionally called the ``resonance widths'')  are expected to reflect the temporal aspects of waves escape from the random medium, and understanding of their properties in various regimes is certainly of considerable interest. Yet, the explicit analytic results on resonance widths statistics remain relatively scarce despite several decades of intensive studies in this direction.

The only class of disordered systems where resonance distribution in the complex plane is presently well understood quantitatively in much detail are wave-chaotic samples in the fully ergodic regime of Quantum Chaos. In such systems, the effects of Anderson localization being negligible, the wave functions extend over the whole available volume of the sample. As a result, a generic/universal description, independent of a particular Hamiltonian, becomes possible as long as the system is probed on time scales longer than the ergodization time $\tau_\text{erg}$ needed for a particle to travel across the system. In this situation, informally called ``zero-dimensional'', the universal features of eigenfunctions, energy levels \cite{Mir2000}, and eventually transport properties \cite{Been11} are controlled by quantum chaos effects and, according to the famous Bohigas-Giannoni-Schmidt conjecture
\cite{BGS1984}, can be efficiently studied using the standard random matrix theory (RMT) framework. The latter is introduced by replacing actual microscopic Hamiltonians describing random medium with random matrices taken from classical Gaussian $\beta$-ensembles, with $\beta=2$ characterising systems with broken (class A) and $\beta=1$ with preserved (class AI) time-reversal invariance.

Various universal aspects of  statistics of the scattering matrix $S(E)$  in the ergodic regime has been be successfully addressed in RMT framework following the pioneering paper \cite{VWZ85}, see Refs.\ \cite{Fyo05_rev,Been11,FSav11,Schomerus2015,StockKuhl_rev_2022} for reviews of that activity. In particular, various questions concerning statistics of resonance poles, and related objects like e.g. associated residues and S-matrix zeroes, has been thoroughly studied along these lines theoretically \cite{SokZel89,Haake1992,Fyo96,FyoSom97,SavSok97,FyoKhor99,Som99,scho00,FM2002,FyoSav12,FyoSav15,li_random_2017,FSK_zeroes_2017,Fyo_zeroes_2019,OsmanFyodorov2020}, and some of theoretical predictions subsequently tested in experiments in wave chaotic scattering from microwave or optical cavities  \cite{kuhl_resonance_2008,Frata12,Frata14,Gros_etal_shifts_2014,Davy_Genack_2018,DavyGenack19} and graphs \cite{CAF_2021_B}, or accurate numerical simulations in realistic models \cite{Kott_res_00,Kolov02,Celardo07}. It is appropriate also to mention  an alternative line of research going back to \cite{GR89}  and relying upon semiclassical methods to address resonances in quantum chaotic systems \cite{LSZ2003,ST2004,KNS2008,SWM2009,Novaes13}.
Note also a growing interest in studying the resonance statistics in such and related systems by rigorous  mathematical methods, both in semiclassical \cite{NZ2007,NSZ2014} and random matrix framework \cite{Kozhan07,DubErd2021,MSTS2021,FKP2022}.

The standard microscopic way of modelling a generic single-particle quantum propagation in a $d$-dimensional random medium filling in a domain ${\cal D}\in \mathbb{R}^d$ amounts to using the Hamiltonian
\be \label{Andcont}
H=-\frac{\hbar^2}{2m}\frac{d^2}{d{\bf r}^2}+V({\bf r})
\ee
 with the appropriate (e.g. Dirichlet) conditions at the boundary of ${\cal D}$. Such Hamiltonian combines the kinetic-energy Laplacian $-\frac{d^2}{d{\bf r}^2}$\footnote{We will set $\hbar=1=2m$ henceforth for brevity.} and a short-ranged correlated  random potential $V({\bf r})$, ${\bf r}\in \mathbb{R}^d$, which in the simplest choice is Gaussian $\delta$-correlated: $\langle V({\bf r}) \rangle=0, \, \langle V({\bf r}) V({\bf r}')\rangle=\frac{1}{2\pi \nu \tau}\delta({\bf r}-{\bf r}')$, with $\nu$ being mean density of energy levels per unit volume and $\tau$ standing for the mean-free time. This defines the main classical characteristic of the disorder: the microscopic diffusion coefficient
 at a fixed energy $E_F$ given by $D\sim E_F\tau$.
One may also include the magnetic field effects in the Hamiltonian  by employing the standard substitution $\frac{d}{d{\bf r}} \to \frac{d}{d{\bf r}}-i\frac{e}{c}{\bf A}$, with ${\bf A}$ being the corresponding vector potential which serves to break the time-reversal invariance of the system. Alternatively, one may use a tight-binding analogue of \eqref{Andcont} on a lattice ${\bf r}\in \mathbf{\Lambda}\subset\mathbb{Z}^d$, represented via
\be\label{tightbinding}
H=\sum_{{\bf r}\in \mathbf{\Lambda}}\,V({\bf r})\left|{\bf r}\right\rangle\left\langle{\bf r}\right|+\sum_{{\bf r}\sim {\bf r}'}\left(t_{{\bf r}{\bf r}'}\left.|{\bf r}\right\rangle\left\langle {\bf r}'\right|+t_{{\bf r}'{\bf r}}\left|{\bf r}'\right\rangle\left\langle {\bf r}\right|\right),
\ee
where the second sum is over nearest neighbours on the lattice, and hopping parameters are assumed to satisfy $t^*_{{\bf r}{\bf r}'}=t_{{\bf r}'{\bf r}}$ to ensure the Hermiticity of the Hamiltonian: $H=H^{\dagger}$, where we use $t^*$ to denote complex conjugation of $t$ and $H^{\dagger}$ for Hermitian conjugation of $H$. Note that the form
\eqref{tightbinding} can be used also for modelling a quantum particle
motion on any graph ${\bf r}\in \mathfrak{G}$ , with corresponding $t_{{\bf r}{\bf r}'}$ being the elements of the adjacency matrix of the graph $\mathfrak{G}$.
The disordered tight-binding model is frequently called in the literature the Anderson model.

As is well known, the major single-particle wave-interference effect due to potential disorder  is the Anderson localization phenomenon which
ensures that generically, for big enough variance in the on-site potential $V({\bf r})$,  the eigenfunctions of Hamiltonians \eqref{Andcont} or \eqref{tightbinding}  around a given energy become localized, with exponentially decaying profile of a characteristic extent $\xi$.
As a result, the classical single-particle diffusive dynamics, characterised in a random medium by a microscopic diffusion constant $D$,
is stopped at distances exceeding $\xi$. At the level of wave scattering this, in particular, implies that  the waves incident from outside on a semi-infinite sample of such medium can penetrate its bulk only to the finite length of the order of $\xi$, with wave intensity exponentially decaying at larger distances. In the low-dimensional random medium with $d\le 2$ the localization length $\xi$ remains finite for any degree of disorder. In higher dimensions the picture is qualitatively different:  generically
  the localization length $\xi$ increases  with decreasing variance of the potential and eventually diverges at a critical disorder characterizing the so-called Anderson (de)localization transition. With further decrease of the disorder the eigenfunctions become extended, i.e. filling in the whole sample volume randomly but uniformly, which restores a diffusive spread inside the medium and ensures transmission of a finite portion of incident wave intensity through the sample.
  The subject of the Anderson localization, especially associated anomalous quantum diffusion and multifractality of eigenfunctions in the vicinity of the delocalization transition, remain the topic of vigorous research activity for several decades, see the review \cite{EversMirlinRev}.

 In such a setting the resonance widths statistics in systems with strong localization effects has been most consistently addressed in the simplest case of wave incident on $d=1$ disordered chain of finite extent $L$, by combination of analytical and numerical methods over the last two decades  \cite{TF_1dres_00,TG00,KS_res_08,1d_Fein09_res,1d_Fein11_res,GS_res_12,1d_Izrailev_res}, see also a review on mathematically rigorous approach to the problem \cite{Klopp16}. The emergent consensus of those works points towards existence of the tail  in the probability density $\rho(\Gamma)\sim 1/\Gamma$ for small $\Gamma\propto e^{-L/\xi}$ which is easy to explain
 by associating imaginary parts $\Gamma$ with inverse time of travel to the boundary of the sample,  and invoking exponential localization of eigenfunctions with characteristic length $\xi$.

 Another result which is worth particular mentioning pertains to {\it quasi}-1D systems, such e.g. as wires of a finite transverse diameter, big enough to support many transverse modes of wave propagation, but still small in comparison with the localization length due to disorder. In contrast to strictly 1D chains, where the localization length $\xi$ and the mean-free path $l$ are typically of the same order, in quasi-1D systems $l\ll \xi$ \cite{EL83}, which
  makes the classical diffusion along the wire possible on scales $l\ll L\lesssim \xi$. The localization in quasi-1D systems attracted a lot of attention due to its relevance to
  the problem of quantum chaos \cite{Izrailev90}, and related interest in a special class of random matrices known as Random Banded Matrices (RBM), see
  \cite{Cas90_RBM,FyoMirRBM91,FyoMirRBM94} for a discussion at the level of Theoretical Physics and \cite{Bourgade_RBM18} for rigorous mathematical results.

   The distribution of resonance widths in quasi-1D setting was first considered in quantum chaos context in \cite{BGS91},
    where another powerlaw tail $\rho(\Gamma)\propto \Gamma^{-3/2}$  in the probability density has been numerically discovered and qualitatively explained  using (semi)classical arguments associating the resonance width with inverse time to reach the boundary via classical diffusion.  To the best of our knowledge, no controllable {\it ab initio} analytic derivation of such a behaviour was reported in the literature starting from any microscopic model Hamiltonian, though the same tail has been later observed in several other models, see e.g. \cite{Skivtiggelen06,BG01}.  One of the original motivations for the present work was to provide such a derivation on equal footing for both tails: $\rho(\Gamma)\sim \Gamma^{-1}$ and  $\rho(\Gamma)\sim \Gamma^{-3/2}$, starting from a well-defined microscopic model. We will see that this is indeed possible, with the derivation simultaneously clarifying limits of applicability for the above semiclassical picture of escape to be operative, and pointing towards yet another universal powerlaw tail: $\rho(\Gamma)\sim \Gamma^{-2}$ for the larger range of width $\Gamma$.

   Results for statistics of resonances in higher dimensional disordered samples $d>1$, in particularly close to the Anderson transition regime,  remain largely numerical, see e.g. \cite{Pinheiro04,Kott05,Skivtiggelen06,Gaspard_Sparenberg_2022,scattPRBM} for a discussion and further references.

\begin{figure}
\centering
\includegraphics[width=40mm]{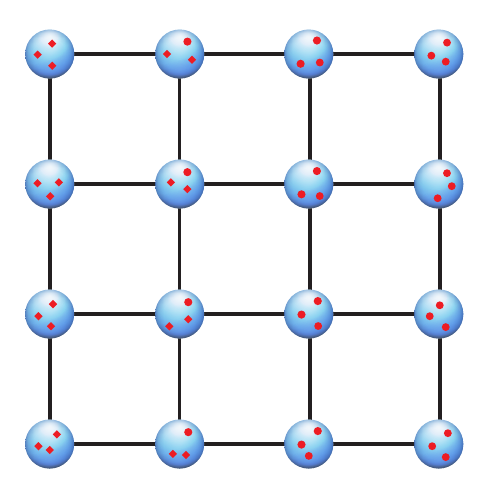}
\caption{A sketch of “granulated
metal” model. Granulas are situated in the sites of a regular lattice, and are assumed to be macroscopically identical but microscopically
different, each one containing its own random configuration of impurities. In the appropriate energy range wavefunctions inside each isolated granula are assumed to be fully delocalized/ergodic, and energy levels described by Wigner-Dyson random matrix staistics. Tunneling between neighboring granulas ensures a possibility of wavefunction spreading accross the lattice sites. Depending on parameters this may either lead to a delocalization at macroscopic distances, or to Anderson localization with finite localization length.}
\label{F:granulated}
\end{figure}

As is commonly accepted nowadays, one of the most powerful and systematic approaches to addressing universal features of wave propagation in a disordered medium, including Anderson localization phenomena, is based on exploiting the framework of the supersymmetric nonlinear sigma model developed in the seminal works by Efetov \cite{Efe_book} building on earlier ideas of Wegner and collaborators \cite{Wegner79,Schafer_Wegner}. In its discrete version the model describes interaction between supermatrices (i.e.\ matrices with Grassmann/anticommuting/ fermionic and ordinary/commuting/bosonic entries)
$Q({\bf r})$, satisfying constraints $Q^2({\bf r})=1$ and $\Str Q({\bf r}) = 0$ for
every site located on a lattice ${\bf r}\in \mathbf{\Lambda}$
 (or more generally, a graph ${\bf r}\in \mathfrak{G}$). Here $\Str$ stands for the standard generalization of the trace to the case of supermatrices,
carrying opposite signs for fermionic and bosonic degrees of freedom, with the size of supermatrices involved depending on underlying symmetries of the Hamiltonian $H$.
In the present paper we are going to consider only the simplest case of the Hamiltonians with fully broken time-reversal symmetry, denoted in the standard
nomenclature as class~A with Dyson parameter $\beta=2$, where the supermatrices are of the size $4\times 4$.
Interaction between such supermatrices at different sites ${\bf r}$ is characterized by the action
\be\label{nlsigmalattice}
S[Q]=\frac{\alpha}{2}\sum_{{\bf r}\sim {\bf r}'}\Str\left[Q({\bf r})Q({\bf r}')\right]+\frac{\pi\eta}{\delta}\sum_{{\bf r}\in \mathbf{\Lambda}}\Str\left[Q({\bf r})\Lambda\right]
\ee
where  the supermatrix $\Lambda$ is diagonal and
 in our case of the broken time-reversal symmetry given by $\Lambda=\diag(1,1,-1,-1)$ in a certain basis.
The model (\ref{nlsigmalattice}) can be derived in a controlled way
from either the so-called Wegner $n$-orbital version of the
Anderson model \cite{Wegner79,Schafer_Wegner}, or from its “banded matrix”
version \cite{FyoMirRBM91,FyoMirRBM94} (in that instance, even fully rigorously \cite{SchSch19,SchSch21}),
 or, perhaps more physically from the “granulated
metal” model \cite{Efe_book}, sketched in Fig.\ \ref{F:granulated}.

For the latter model, the interaction parameter
$\alpha$ in the action (\ref{nlsigmalattice}) is proportional to tunneling
between coupled granulas, and is related to the microscopic
diffusion constant $D$ as $\alpha = \pi\nu D a^{d-2}$, where $a$ stands for the granula size. The parameter $\eta>0$ is the small
imaginary increment of the energy necessary to make
the (retarded) Green’s function of the underlying microscopic
model $G({\bf r},{\bf r}',E+i\eta) := \left\langle{\bf r}\right|(E+i\eta-H)^{-1}\left|{\bf r}'\right\rangle$  well-defined,
and physically describes the spatially-uniform
absorption rate inside the medium.
It is normalized by $\delta = 1/\nu a^d$ where $\nu$ is the density of states (DOS)  per unit volume. Treatment
of classical waves along the same lines can be found, for example,
in Ref. \cite{Tian2013_rev}.

Although working with the lattice version Eq.\ (\ref{nlsigmalattice}) of the nonlinear $\sigma$-model is quite natural, when describing a sample of continuous disordered medium filling in a finite domain ${\cal D}\in\mathbb{R}^d$
of  a characteristic size  $L$ it is more convenient to  consider the corresponding continuum limit
 governed by the action
\begin{equation}\label{nlsigmacont}
{\cal S}[Q]
=
-\pi\nu\int_{\cal D} \Str\left[\frac{D}{4}(\nabla Q)^2 - \eta\, Q\Lambda\right] d^d{\bf r} ,
\end{equation}
where $\nu$ is the density of states per unit volume defining the mean level spacing $\Delta=(\nu L^d)^{-1}$ in a sample of size $L$, and $\eta$ is the uniform absorption rate, which can be formally identified with the (twice) imaginary frequency. The action Eq.\ (\ref{nlsigmacont}) can be obtained from the lattice version Eq.\ (\ref{nlsigmalattice}) by allowing the lattice constant $a\to 0$ and
 formally identifying $\sum_{{\bf r}\in \mathbf{\Lambda}}=a^{-d}\int_{\cal D}d^d{\bf r}$ in such a limit, which e.g. gives for the interacting term
\begin{gather*}
\sum_{{\bf r}\sim {\bf r}'}\Str\left[Q({\bf r})Q({\bf r}')\right]=-\frac{1}{2}\sum_{{\bf r}\sim {\bf r}'}\Str\left[Q({\bf r})-Q({\bf r}')\right]^2\approx
-\frac{a^2}{2} \sum_{{\bf r}\in \mathbf{\Lambda}}\Str \left[\nabla Q({\bf r})\right]^2
\\{}
= -\frac{a^{2-d}}{2}\int_{\cal D} \Str \left[\nabla Q({\bf r})\right]^2\,{d^d{\bf r}}
\end{gather*}

Note that all interesting wave interference phenomena (in particular, related to the Anderson localization)
require presence of both advanced and retarded Green's function for their description, and at the level of nonlinear $\sigma$-model this fact is reflected in
the structure of the supermatrices in their retarded-advanced basis, most prominently in  $\Lambda$ having opposite sign in the upper-left and bottom-right blocks.
Calculation of various quantities characterizing universal
features of statistics of eigenfunctions and energy levels in disordered media can be reduced to studying expectations
of various combinations of different supermatrices $Q({\bf r})$ over the
weight $\exp\left(-S\left[Q\right]\right)$ \cite{Efe_book,MirPhysRep00}. In particular,  one of the central objects of the theory, the
so called “order parameter function” (OPF) introduced
originally in Ref.\ \cite{Zirn86} is defined as
\begin{equation}\label{OPF-def}
F_{\bf r}(Q;\eta)=\int\prod_{{\bf r}'\ne {\bf r}}{\cal D}\mu\left(Q_{\bf r'}\right) \exp(-{\cal S}[Q]) ,
\end{equation}
where ${\cal D}\mu\left(Q_{\bf r'}\right)$ is the standard volume element on the manifold of $Q$ matrices.
Due to global symmetries of the action (\ref{nlsigmalattice}), the OPF  can be shown to actually depend only on a few real Cartan variables parametrizing $Q$ matrices. In particular, for $\beta=2$ one has $F_{\bf r}(Q;\eta):={\cal F}(\lambda,\lambda_1;\eta)$, with $\lambda\in[-1,1]$ and $\lambda_1\in[1,\infty]$ being the compact (fermionic) and non-compact (bosonic) coordinates, respectively (we omitted spatial dependence on ${\bf r}$ for brevity).

\begin{figure}
\centering
\includegraphics[width=55mm]{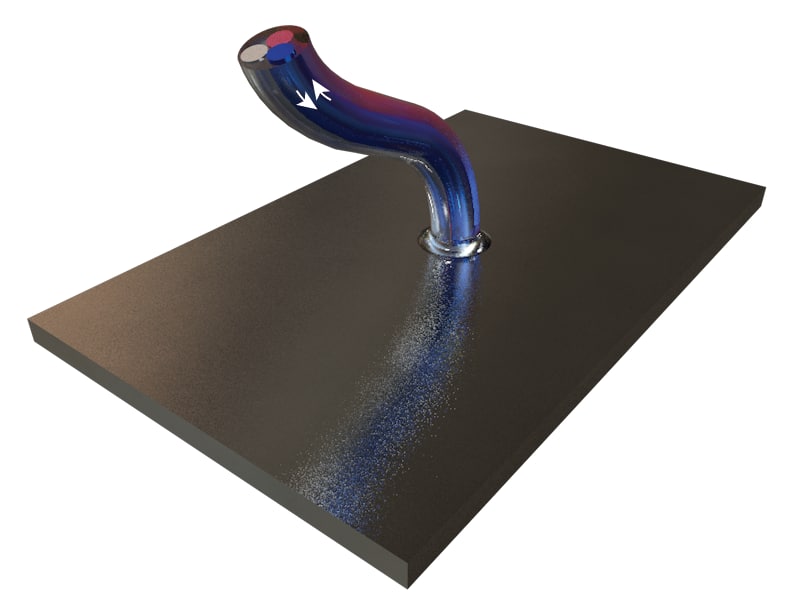}
\caption{A wave-guide with $M$ propagating channels coupled to a disordered medium.}
\label{F:setup}
\end{figure}

In the present paper we first aim at showing how
to express in terms of the OPF Eq.~(\ref{OPF-def}) the density of the complex $S$-matrix poles for the problem of a single multi-channel lead scattering from a disordered
medium, see the Fig.~\ref{F:setup}, and then to study the ensuing resonance widths distribution in various parameter regimes.

Note that the OPF characterizes the {\it closed} system which conserves the number of particles, whereas allowing particles/waves at a given energy to be sent via the lead to the random medium and then collecting the reflected waves renders the medium {\it open}.
We will see that a possibility to relate properties of open systems (in particular, the density of resonance poles in the complex plane) to the characteristic of its closed counterpart is eventually based on the assumption of ``locality" of the lead, whose transverse extent is assumed to be much smaller
than the mean-free path in the disordered medium, making the coupling effectively point-wise at the level of $\sigma$-model description. Still, even such point-wise lead
may support arbitrary many propagation channels, though we will be always assuming $M$ remaining negligible to the number of sites in the underlying lattice~$\mathbf{\Lambda}$.

\subsection{Overview and structure of the paper}
Consider a model of random medium consisting of macroscopically identical but microscopically
different metallic granules associated to each site ${\bf r}$ in a
lattice or graph, and denote the associated Hamiltonian as $H$ which for simplicity we will consider belonging to
the symmetry class $\beta=2$. Assume that $H$ is characterized
by $N$ energy levels $E_n$ with mean level spacing
$\Delta = 1/[N\nu(E)]$, where $\nu(E)$ stands for the mean DOS around point $E$ of the spectrum.

Consider now attaching a lead/waveguide supporting $M$ propagating channels at a given energy $E$ to one of such
granula and use it to send particles/waves
at this energy to the random medium and collect the reflected waves. As a result of the coupling to the lead each real energy level $E_n$ of a closed system is converted to a complex resonance pole ${\cal E}_n=E_n-i\Gamma_n$. The value for the mean (i.e. disorder-averaged, denoted by angular brackets $\left\langle ...\right\rangle$) density of such poles in the complex plane is given by
\begin{equation}\label{resdendef}
 \rho_E(y):=\Delta\,
 \biggl\langle
 \sum_{n=1}^N \delta(E-E_n) \delta\left(y-2\pi \Gamma_n/\Delta\right)
 \biggr\rangle ,
\end{equation}
where we switched to the rescaled widths $2\pi \Gamma_n/\Delta$. %In what follows, we neglect the energy dependence of the mean level spacing $\Delta$.
Our goal is to develop the theory for evaluating $\rho_E(y)$, assuming $N\to \infty$ at a fixed number of channels $M$, which otherwise remains an arbitrary positive integer.

In pursuit of this goal we will follow the framework of the standard (Heidelberg) model \cite{VWZ85} which we briefly discuss in the beginning of Section \ref{S:2}.
 The couplings between the lead and the disordered medium in such framework are characterized via the effective  coupling parameters $g_a=2/T_a-1\geq 1$ ($a=1,\ldots, M$) related to the Landauer transparencies $T_a=1-\left|\left<S_{aa}\right>\right|^2$ which characterize the $M\times M$ ensemble- (or energy-) averaged scattering matrix $S(E)$.  For simplicity and clarity of the presentation we will consider in the paper explicitly only the case of ``equivalent channels'' $g_a:=g$, though extension to non-equivalent channels of different coupling strength is not difficult to include.

Our main technical achievement amounts to explicitly relating the resonance density $\rho_E(y)$ to the Order Parameter Function  $F_{\bf r}(Q;\eta):={\cal F}(\lambda,\lambda_1;\eta)$ defined in Eq.\ (\ref{OPF-def}) for the medium with broken time-reversal symmetry (class A, $\beta=2$). Namely, replace in the OPF ${\cal F}(\lambda,\lambda_1;\eta)$ the absorption/imaginary frequency parameter $\eta$ with the  resonance width as $\eta\to \Gamma=y\Delta/2\pi$ and
define the function
\begin{subequations}
\label{resdenmainM}
\begin{equation}\label{resdenmainM=anyB}
  \Phi(\lambda,\lambda_1;y)
  :=
  \frac{\partial^2}{\partial y^2}
  \frac{\partial^{M-1}}{\partial\lambda_1^{M-1}}
  \frac{{\cal F}(\lambda,\lambda_1;y\Delta/2\pi)}{(\lambda_1-\lambda)^2}
\end{equation}
 by taking derivatives of the OPF with respect to the non-compact coordinate $\lambda_1$ and with respect to $y$.
 Then the resonance density is given by the following expression:
\begin{equation}\label{resdenmainM=anyA_text}
  \rho_E(y)
  =
  \frac{(-1)^{M-1}}{2(M-1)!}
  \int_{-1}^1 (g-\lambda)^M\,
  \Phi(\lambda,g;y)\,d\lambda ,
\end{equation}
\end{subequations}
valid for a waveguide supporting $M$ identical scattering channels, each with the same coupling constant $g_a=g$.
Note that in the above equation the coordinate $\lambda_1$ gets pinned, $\lambda_1\to g$.  It therefore remains to integrate only over the compact coordinate $\lambda$.
Thus the knowledge of the OPF for a {\it closed} system as a function of the finite uniform absorption rate $\eta$ fully determines the density of resonances in the complex plane for its open counterpart.

A detailed derivation of Eqs.\ (\ref{resdenmainM}) is provided in Section \ref{S:2.2}. Here we only mention that it combines two essential ingredients:
\begin{itemize}
\item
The recently discovered relation \cite{CAF_2021_B} between the density $\rho_E(y)$ and a complex-valued generalization  \cite{CAF2021A} of the standard Wigner time delay \cite{Wigner55}.
\item
A relation between the OPF and the mean density of complex eigenvalues $K_c, \, c=1,\ldots, M$
of the so-called Wigner $K$-matrix. The latter matrix is related to the scattering $S$-matrix via the matrix Cayley transform as $S=(1-iK)/(1+iK)$ and in the context of electromagnetic wave scattering also has a meaning of the impedance matrix, see e.g. \cite{Hemmadi05}). Note that the relation between the OPF and the mean density of Wigner matrix eigenvalues generalizes a well-known relation between the OPF and the joint probability density  ${\cal P}(u,v)$ of real $u$ and complex $v$ parts of the local Green's function $G({\bf r},{\bf r},E+i\eta)=u-iv$  \cite{MirFyo1994,savfyodsom05}.
\end{itemize}

Equations (\ref{resdenmainM}) reduce the analysis of the mean density of the resonance poles in the framework of nonlinear $\sigma$-model to exploiting the knowledge of the functional dependence of the OPF on its parameters.
In Section \ref{S:3} of the paper we analyse various cases where such knowledge is available. We will mainly concentrate on particular on the ``perfect coupling'' limit when
$g=1$, corresponding to the vanishing mean $\corr{S_{aa}(E)}=0$, and accounting for the fully open channels in the leads. The latter condition physically corresponds to the absence of the so-called fast ``direct reflection'' processes, so that all the incoming flux penetrates the medium and participates in the formation of long-living resonances, see Refs.\ \cite{Fyo05_rev,FSav11} for a discussion and further references. Finally in Section \ref{S:4} we summarize the outcome of our analysis and briefly discuss the results.

\section{Derivation of the main relation for the mean resonance density,  Eq.\ (\ref{resdenmainM})}
\label{S:2}

\subsection{A short reminder on the Heidelberg model}

\begin{figure}
\centering
\includegraphics[width=70mm]{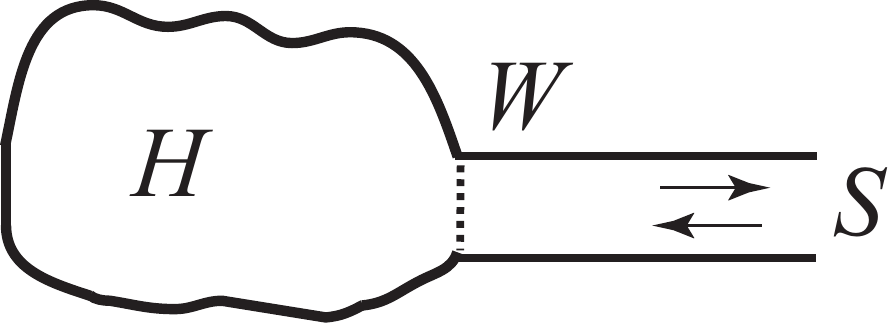}.
\caption{A sketch of a chaotic wave scattering from a region schematically represented by a cavity and assumed to contain a random medium inside.
 An operator governing wave dynamics in such a cavity
decoupled from the channels is assumed to be effectively described by a large random matrix $H$. An infinite lead is assumed to support $M$ propagating channels
in the considered energy range, and is coupled to the cavity region by a matrix/operator $W$. The  ensuing $M\times M$ unitary scattering matrix $S$
can be related to $H$ and $W$ in the framework of the Heidelberg approach, and is given by Eq.\ (\ref{1}).}
\label{F:sketch}
\end{figure}

An efficient way of describing scattering of classical or quantum waves, especially convenient in the case of random medium Hamiltonians whose statistical properties can be mapped to a non-linear sigma-model,
has been formulated in Ref.\ \cite{VWZ85}, see e.g.\ \cite{FyoSom97} for more detail. Within such a framework, which is frequently called in the literature the ``Heidelberg model'',  one constructs  the unitary (in the absence
of absorption, $\eta=0$) $M\times M$ energy-dependent scattering matrix $S(E)$  describing scattering of waves incident on  a random medium and then exiting it via $M$ open scattering channels, which in case of electromagnetic waves are associated with $M$ antennae, see the sketch in Fig.\ \ref{F:sketch}.

Unitarity reflects the flux conservation: the vectors ${\bf a}=(a_1,\ldots,a_M)$ of incoming and ${\bf b}=(b_1,\ldots,b_M)$ of outgoing
amplitudes are linearly related via ${\bf b}=S{\bf a}$ and have the same norm.
The medium is considered to be confined to a spatial domain ${\cal D}$ and described by a self-adjoint Hamiltonian $H$, e.g.\ Eq.\ (\ref{Andcont}) or its tight-binding analogue Eq.~(\ref{tightbinding}), as discussed in the Introduction.
The latter case is convenient as allows one to think of such a Hamiltonian as described by a large $N\times N$ random matrix $H$, but a similar effective description
 is equally valid for the continuum version, in an appropriate energy range and after due modifications. The relation between  $S(E)$ and $H$  is then provided by the following formula
\begin{equation}\label{1}
S(E)=\frac{{\bf 1}-iK}{{\bf 1}+iK}, \qquad
K=W^{\dagger}\frac{1}{E-H}W
\end{equation}
and columns $W_a$ ($a=1,\ldots, M$) of an $N\times M$ matrix $W$ of  {\it coupling amplitudes}\/ to $M$ open scattering channels can be taken as fixed vectors satisfying
the {\it orthogonality condition}:
\be \label{orth}
\sum_{i=1}^N W^*_{ai}W_{bi}=\gamma_a\delta_{ab},
\ee
 with $\gamma_a>0$ $\forall a=1,\ldots, M $  determining the ``bare'' strength of coupling of a given channel to the scattering system.
The orthogonality condition ensures that the ensemble-averaged scattering
matrix can be assumed to be diagonal\footnote{In fact, there is no reduction of generality in assuming $S$-matrix diagonality, see e.g. \cite{NW85}, and no much gain in dispensing with the channel orthogonality, see e.g. appendix A in \cite{FF20}.} as  long as the scattering system allows effective description in the nonlinear sigma-model framework:
\be \label{diagS}\corr{S(E)}=\diag
 (\corr{S_1(E)},\ldots,\corr{S_M(E)}), \quad \corr{S_a(E)}=\frac{1-i\gamma_a\corr{G}}{1+i\gamma_a\corr{G}},
 \ee
  where we introduced $\corr{G}:=\lim_{\eta\to 0} \corr {G({\bf r},{\bf r},E+i\eta)}=\corr{u}-i\pi\nu(E)$, with $\nu(E)$ being the mean density of states in the disordered medium.
  Writing $\corr{G}=|\corr{G}|e^{-i\alpha}$ and defining $\tilde{\gamma}_a:=\gamma_a |\corr{G}|$ one finds
 \be
  |\corr{S}_a|^2=\frac{1-2\sin{\alpha}\,\tilde{\gamma}_a+\tilde{\gamma}_a^2}{1+2\sin{\alpha}\,\tilde{\gamma}_a+\tilde{\gamma}_a^2}
 \ee
 implying
 \be
  g_a=\frac{1+|\corr{S_a(E)}|^2}{1-|\corr{S_a(E)}|^2}=\frac{1}{2\sin{\alpha}}\left(\tilde{\gamma}_a+\frac{1}{\tilde{\gamma}_a}\right)\ge 1.
\ee
The set of parameters $g_a, \, a=1,\ldots,M$ provides the complete description of coupling of the medium to scattering channels in the universal regime, with
 the ``perfect coupling'' value $g_a=1$ (happening when $\sin{\alpha}=1$ and $\tilde{\gamma}_a=1$) corresponding to $|\corr{S_a}|=0$. The latter condition physically
 implies absence of short-time (also known as ``direct'') scattering processes at the channel $a$ entrance: all the incoming flux penetrates inside the medium and participates in formation of long-living resonant structures. This situation is thus most interesting from theoretical point of view, and is frequently described by most elegant formulas.

 Equivalently, entries $S_{ab}(E)$ of the scattering matrix given by (\ref{1}) can be rewritten as
\begin{equation}\label{2}
S_{ab}(E)=\delta_{ab}-2i\sum_{ij} W^*_{ai}\left[\frac{1}{E-{\cal H}_\text{eff}}\right]_{ij}W_{jb},
\end{equation}
with an effective non-Hermitian Hamiltonian
\begin{equation}\label{3}
 {\cal H}_\text{eff}=H-i\Gamma, \, \quad \, \Gamma=WW^{\dagger}\ge 0
\end{equation}
whose $N$ complex eigenvalues ${\cal E}_n=E_n-i\Gamma_n$ provide poles of the scattering matrix in the complex energy plane, commonly referred to as the {\it resonances}. The goal of this paper is precisely to describe density of these poles in the complex plane.

\subsection{Derivation of Eq.\ (\ref{resdenmainM})}
\label{S:2.2}
  We recall that the scattering matrix $S(E)$ as defined in (\ref{1}) or (\ref{2}) is unitary as long as the energy parameter $E$ is kept real.
   For our goals it is however expedient to introduce a finite absorption rate $\eta>0$ inside
the scattering region, which formally amounts to replacing $E\to E+i\eta$ in the definition of the $S$-matrix. Such replacement immediately renders $S$-matrix subunitary,
reflecting the induced irretrievable flux absorption inside the medium. Although originally introducing the absorption parameter $\eta$ into the model has been motivated by practical needs to describe realistic experiments where losses can never be avoided, recently it was realized \cite{CAF_2021_B} that such parameter may be used for counting
$S$-matrix poles, both theoretically and experimentally.  Note also that the substitution $E\to E+i\eta$ implies that the $M\times M$ matrix $K$ defined in Eq.\ (\ref{1}) becomes now non-Hermitian: $K^{\dagger}\ne K$, and its eigenvalues $K_a$ become in general complex-valued.

Given such subunitary  $M$-channel scattering matrix $S(E+i\eta)$, one now may follow \cite{CAF2021A} and define the complex-valued quantity
\be
\tau_W(E,\eta)=-i\frac{\partial}{\partial E} \log\det{S(E+i\eta)}\equiv \frac{\partial}{\partial \eta} \log\det{S(E+i\eta)},
\ee
The quantity $\tau_W(E,\eta)$ is obviously a complex-valued generalization of  the standard Wigner time delay, and reduces to it for $\eta\to 0$ [4]. As has been found in \cite{CAF_2021_B} the following relation exists between
the mean resonance density $\rho_E(y)$  defined in Eq.\ (\ref{resdendef}) and the above-defined $\tau_W(E,\eta)$:
\begin{equation}\label{relation1}
\Re \left\langle \tau_W(E,\eta) \right\rangle = \frac{2\pi}{\Delta}\int_{\tilde{\eta}}^{\infty}\rho_E(y)\,dy
\end{equation}
where $\tilde{\eta}:=\frac{2\pi \eta}{\Delta}$, and $\Delta$ is the associated mean level spacing. Using this relation one then arrives to the following representation for the mean resonance density:
\begin{equation}\label{relation2}
\rho_E(y)= -\frac{\partial^2}{\partial y^2}\left\langle \log{\left|\det S\left(E+iy\frac{\Delta}{2\pi}\right)\right|} \right\rangle
\end{equation}
We see that the absorption $\eta$ get replaced by the resonance widths $\Gamma =y\frac{\Delta}{2\pi}>0$.
Now, using Eq.\ (\ref{1}) we further see that
\be\label{relK}\nonumber
\log{\left|\det S\left(E+iy\frac{\Delta}{2\pi}\right)\right|}
=\sum_{a=1}^M\log{\left|\frac{1-iK_a}{1+iK_a}\right|}
\ee
\be
=\int \rho_M\left(\Re K, \Im K; y\frac{\Delta}{2\pi}\right)\log{\left|\frac{1-iK}{1+iK}\right|}\,d^2K,
\ee
where we introduced the density
\begin{equation}\label{denK}
 \rho_M(\Re K,\Im K;\eta)=\sum_{a=1}^M\delta\left(\Re K-\Re K_a\right) \delta\left(\Im K-\Im K_a\right)
\end{equation}
of the complex eigenvalues $K_a=\Re K_a-i\Im K_a$, $\forall a=1,\ldots, M$ of the (non-Hermitian) $K$-matrix  defined via \eqref{1} at the complex energy $E+i\eta$.
Here we explicitly indicate the dependence on the parameter $\eta$,  recalling again that $\eta$ should be eventually replaced by the resonance widths $\Gamma =y\frac{\Delta}{2\pi}>0$. Thus, to evaluate the mean resonance density $\rho_E(y)$ one needs to know the ensemble-averaged value $\left\langle \rho_M(\Re K, \Im K;\eta)\right\rangle$ for a given $\eta$.  The consideration so far has been pretty general and can be applied to any microscopic model of a disordered medium described by a Hamiltonian matrix $H$, assuming validity of the Heidelberg construction (\ref{1}) for the $S$-matrix, without any reference to a nonlinear $\sigma$-model.

The power of nonlinear $\sigma$-model description lies in our ability to provide an explicit representation of the mean density  $\left\langle \rho_M(\Re K,\Im K;\eta)\right\rangle$ of $K$-matrix eigenvalue in terms of the Order Parameter Function ${\cal F}(\lambda,\lambda_1;\eta)$. For simplicity and clarity of the ensuing formulas we consider below only the so-called case of equivalent channels, with $\gamma_a=\gamma$ for all $a=1,\ldots,M$, and further
tacitly assume that $\corr{G}=-i$ so that $\gamma=\tilde{\gamma}$. The latter condition is well-known to hold, for example,  at the middle of the semicircular spectrum of random matrix Hamiltonians representing Wegner gauge-invariant $n$-orbital model. In such a case one can write $K_a=\gamma G_a$, with $G_a=u_a-iv_a$, and then show that the associated mean density
\begin{equation}\label{denG}
\rho_M(u,v;\eta):=\left\langle \sum_{a=1}^M\delta\left(u-u_a\right) \delta\left(v- v_a\right)\right\rangle
\end{equation}
 can be represented as
\begin{equation}\label{jpd}
\rho_M(u,v;\eta)=\frac{P_0(x)}{4\pi v^2},
\quad P_0(x)=\frac{d}{dx}(x^2-1)\frac{d}{dx}\phi_M(x),
\end{equation}
where we denoted $x=\frac{u^2+v^2+1}{2v}>1$ and defined
\be\label{jpda}
\phi_M(x)=\frac{(-1)^{M-1}}{(M-1)!}
\int_{-1}^1(x-\lambda)^{M}\frac{\partial^{M-1}}{\partial x^{M-1}}\left[\frac{{\cal F}(\lambda,x;\eta)}{(x-\lambda)^2}\right]\,d\lambda.
\ee
For $M=1$ the expression above is well-known, see \cite{MirFyo1994}, and provides the joint probability density ${\cal P}(u,v)$ of real $u$ and complex $v$ parts of the diagonal element $G({\bf r},{\bf r},E+i\eta)=u-iv$ of the Green's function (to which the $K$-matrix is proportional in $M=1$ case) in terms of the OPF.
 For a general integer number of channels $M\ge 1$ the expressions (\ref{jpd})-(\ref{jpda}) can be derived by generalizing the approach suggested in \cite{savfyodsom05}, and the corresponding calculation will be published elsewhere \cite{Fyo_unpub}. Note that especially simple form of the relation between $\phi_M(x)$ and OPF ${\cal F}(\lambda,x;\eta)$ in (\ref{jpda}) is specific
for systems with broken time reversal invariance (class A, $\beta=2$). For preserved time-reversal invariance (class AI, $\beta=1$),
the relation has exactly the same functional form, but explicit expression for $\phi_M(x)$  in terms of the corresponding  OPF is considerably more complicated \cite{savfyodsom05,Fyo_unpub}.

Having (\ref{jpd}) at our disposal, the calculation of the resonance density starts with rewriting in the integrand of (\ref{relK})
\begin{equation}\label{Smat1}
\left|\frac{1-iK}{1+iK}\right|=\left|\frac{1-i\gamma (u-iv)}{1+i\gamma (u-iv)}\right|^2=\frac{\frac{1}{2v}(1-\gamma^2)-\gamma+\gamma^2\,x}{\frac{1}{2v}(1-\gamma^2)+\gamma+\gamma^2\,x}, %\quad x=\frac{u^2+v^2+1}{2v}>1
\end{equation}
which allows further to rewrite (\ref{relation2}) in terms of $\rho_M(u,v;\eta) $ for a fixed absorption $\eta>0$ as
\begin{equation}\label{relation3}
\rho_E(y)= -\frac{1}{2}\frac{\partial^2}{\partial y^2}\int_{-\infty}^{\infty} du \int_0^{\infty} dv \, \rho_M(u,v;\eta)  \log{\frac{\frac{1}{2v}(1-\gamma^2)-\gamma+\gamma^2\,x}{\frac{1}{2v}(1-\gamma^2)+\gamma+\gamma^2\,x}} .
\end{equation}

Substituting (\ref{jpd}) into (\ref{relation3}) it is convenient first to pass from $u^2$ to $x$ and from $v$ to $1/v$
as integration variables, yielding the resonance density in the form
\begin{equation}\label{relation4}
\rho_E(y)= -\frac{1}{4\pi}\frac{\partial^2}{\partial y^2}
\int_{1}^{\infty} dx \, P_0(x) J_{\gamma}(x),
\end{equation}
with $J_{\gamma}(x)$ defined as
\begin{equation}
 J_{\gamma}(x)=\int_{v_{-}}^{v_{+}} \frac{dv}{\sqrt{(v-v_{-})(v_+-v)}}  \log{\frac{\frac{v}{2}(1-\gamma^2)-\gamma+\gamma^2\,x}{\frac{v}{2}(1-\gamma^2)+\gamma+\gamma^2\,x}},
\end{equation}
where we denoted $v_{\pm}=x\pm\sqrt{x^2-1}$.
Using the relation (\ref{jpd}) between $P_0(x)$ and $\phi_M(x)$ and the fact that $J_{\gamma}(x\to \infty)\to 0$ one can use integration by parts yielding
\begin{equation}\label{byparts}
\int_{1}^{\infty} dx \, P_0(x) J_{\gamma}(x)=-\int_{1}^{\infty}\, \tilde{P}_0(x)\frac{\partial}{\partial x}J_{\gamma}(x)\, dx, \quad
\tilde{P}_0(x)= (x^2-1)\frac{d}{dx}\phi_M(x) .
\end{equation}

To evaluate $\frac{\partial}{\partial x}J_{\gamma}(x)$ one first changes $v=v_{-}+(v_{+}-v_{-})\frac{1+\cos{\theta}}{2}, \, \theta\in[0,\pi]$ which after introducing the effective coupling parameter $g=\frac{1}{2}(\gamma+\gamma^{-1})$ so that
$\sqrt{g^2-1}=\frac{1}{2}\left|\gamma-\gamma^{-1}\right|$ allows to bring the integral over $v$ in  (\ref{relation4}) to the form
\begin{equation}\label{relation5}
 J_{\gamma}(x)=\int_{0}^{\pi} d\theta\, \log\,\frac{-1+gx+\sqrt{(g^2-1)(x^2-1)}\cos{\theta}}{1+gx+\sqrt{(g^2-1)(x^2-1)}\cos{\theta}}.
\end{equation}
Now one can use the formula $\int_{0}^{\pi} \frac{d\theta}{a+b\cos{\theta}}=\pi/\sqrt{a^2-b^2}, \, |a|>|b|$ and find after straightforward manipulations that
 \begin{equation}\label{relation6}
\frac{\partial}{\partial x}J_{\gamma}(x)=\frac{2}{x^2-1}
\times
\begin{cases}
  1, & \text{if $x>g$;} \\
  0, & \text{otherwise;}
\end{cases}
%\left\{\begin{array}{cc}1 & \mbox{if} \,\, x>g\\ 0 & \mbox{otherwise}  \end{array}\right.
\end{equation}
which after substituting into (\ref{byparts}) gives
\begin{equation}\label{byparts1}
\int_{1}^{\infty} dx \, P_0(x) J_{\gamma}(x)=-2\int_{g}^{\infty} dx \, \frac{d}{dx} \phi_M(x) = 2 \phi_M(g)
\end{equation}
using ${\cal F}(\lambda,x;\eta)\to 0$ as $x\to \infty$. Equation (\ref{byparts1}) when combined with the definition
of $\phi_M(x)$ in (\ref{jpd}) yields upon substituting to (\ref{relation4}) exactly the main formula Eq.\ (\ref{resdenmainM=anyA_text}) for the mean resonance density
in the complex plane, which we repeat here for convenience:
\begin{equation}\label{resdenmainM=anyA_text2}
  \rho_E(y)
  =
  \frac{(-1)^{M-1}}{2(M-1)!}\int_{-1}^1 (g-\lambda)^M\,
\frac{\partial^{M-1}}{\partial x^{M-1}}
\left[ \frac{\frac{\partial^2}{\partial y^2}{\cal F}(\lambda,x;y)}{(x-\lambda)^2}\right]_{x=g}
d\lambda .
\end{equation}

\section{Analysis of the resonance widths distribution in particular cases}
\label{S:3}

In this section we aim to discuss various particular cases, providing explicit expressions for the mean resonance density  in disordered systems of  spatial dimensions
$0\le d\le 3$, in the range of parameters amenable to analytical treatment. For brevity, we suppress the energy dependence and write $\rho(y)$ instead of $\rho_E(y)$.

\subsection{Zero-dimensional (0D) limit}
The simplest case is that of a disordered medium represented by
a fully ergodic RMT-like system, which physically corresponds to a single metallic granula with fully delocalized eigenstates.
Such systems, conventionally called ``zero-dimensional'', are characterized by the OPF given explicitly in the case of fully broken time-reversal invariance
 by \cite{Efe_book}
\be \label{0DOPF}
{\cal F}(\lambda,\lambda_1;\eta) = e^{-(2\pi\eta/\Delta)(\lambda_1-\lambda)}
\ee
 implying via Eq.\ (\ref{resdenmainM=anyA_text2}) the resonance density
 \be
\label{rho-0D-gen}
  \rho_\text{(0D)}(y)
  =
  \frac{(-1)^{M}y^{M-1}}{(M-1)!}\frac{\partial^M}{\partial y^M}\left(e^{-yg}\frac{\sinh{y}}{y}\right),
\ee
obtained originally by a different method in Ref.\ \cite{{Fyo96}}.
In the perfect coupling limit, this resonance density has a maximum at $y\approx M/2$, with a tail $\rho^\text{(0D)}(y) \approx M/2y^2$ at $y>M/2$.
This formula has been confirmed numerically \cite{Kott_res_00} and, more recently, experimentally \cite{CAF_2021_B} in chaotic wave scattering from microwave graphs  and also validated in accurate numerical simulations of resonances in realistic physical models, e.g. in Stark-Wannier systems subject to time-periodic fields \cite{Kolov02}.

\subsection{Quasi-1D systems: transfer-matrix method}
\begin{figure}
\centering
\includegraphics[width=100mm]{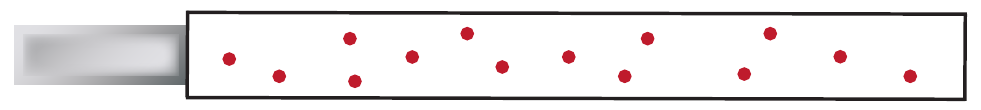}
\caption{A sketch of the “quasi-1D” model. The left part in grey represents an infinite-length ideal lead supporting $M$
propagating modes. The disordered part is of a finite length $L$ and contains finite concentration of random impurities inside.}
\label{F:quasi1d}
\end{figure}

We will consider quasi-1D samples, which physically describe the disordered wires with transverse dimension much larger than the mean-free path (so that the transverse diffusion
is locally possible) but still much smaller than the localization length, so that there is full wavefunction ergodization in the transverse direction.
At one end the sample will be assumed to be closed/isolated (Dirichlet boundary conditions), whereas at the opposite end attached to an infinite-length ideal lead supporting $M$
propagating modes, see the sketch in Fig.\ \ref{F:quasi1d}.

As is well-known since the early work \cite{EL83}, the disordered segment is described in the nonlinear $\sigma$-model framework by the action Eq.\ (\ref{nlsigmacont}),
with the domain ${\cal D}$ specified as the interval $r\in[0,L]$. Such one-dimensional geometry allows one to find the OPF  $F_{\bf r}(Q;\eta)$ via a certain transfer matrix
procedure in terms of a function $\Psi(Q)$
satisfying the following imaginary-time evolution equation
\be
\label{transf-eq}
  \frac{\partial\Psi}{\partial t} = -  \mathcal{H} \Psi,
\ee
with the role of time $t$ played by the appropriately scaled coordinate $r$ along
the system (measured in units of the localization length $\xi$). The evolution equation should be supplemented in the chosen geometry with
the initial condition $\Psi(Q)=1$ at $t=0$ which corresponds physically to the condition of one closed edge of a quasi-1D sample.
 In the unitary symmetry case under consideration the function $\Psi(Q)$ depends only on the compact Cartan variable $\lambda\in[-1,1]$
and non-compact $\lambda_1\in [0,\infty)$, which results in the transfer-matrix Hamiltonian $\cal{H}$ which we will write, see e.g. Eq.\ (9) of \cite{SO07}, as
\be
\label{Ham-lambda}
  \mathcal{H}
  =
  -
  (\lambda_1-\lambda)^2
  \left[
    \frac{\partial}{\partial\lambda_1}
    \frac{\lambda_1^2-1}{(\lambda_1-\lambda)^2}
    \frac{\partial}{\partial\lambda_1}
  +
    \frac{\partial}{\partial\lambda}
    \frac{1-\lambda^2}{(\lambda_1-\lambda)^2}
    \frac{\partial}{\partial\lambda}
  \right]
  + \frac{\Gamma}{\Delta_\xi} (\lambda_1-\lambda) ,
\ee
where for the problem of resonance statistic we replaced $\omega\to 2i\Gamma$, and introduced $\Delta_\xi = (4\pi^2D\nu^2)^{-1}=D/\xi^2$  for the mean level spacing in the samples with length $L$ equal to the localization length $\xi=2\pi\nu D$.

The specific reflection problem we are going to consider explicitly is such that
 the M-channel lead supporting incoming and outgoing waves is attached to one of the ends of the sample, which we
 associate with the coordinate $x=L$,  with the other end at $r=0$ being purely reflecting/isolated, see again Fig.\ \ref{F:quasi1d}.  According to our general formula, Eq.\ (\ref{resdenmainM}) or Eq.\ (\ref{resdenmainM=anyA_text2}),  expression for the resonance density
requires the knowledge of the OPF at the point $r=L$, which in the present scattering geometry simply coincides with the solution $\Psi(\lambda,\lambda_1)$ of the transfer matrix equation (\ref{transf-eq}), explaining the particular convenience of our choice. Note that in a more general situation,   with a lead attached (e.g. transversely to the sample) at an intermediate point $0<r<L$, the corresponding OPF will be related to the product of transfer-matrix solutions $\Psi$ obtained for intervals $[0,r]$ and $[0,L-r]$, correspondingly. This will make the actual calculation technically slightly more involved, though still possible. As we expect all essential features of quasi-1D scattering to be qualitatively similar irrespective of the point of lead attachment, we concentrate on the simplest geometry henceforth.

In general, for arbitrary $t$ the explicit analytical solution of the transfer-matrix equation (\ref{transf-eq}) is not yet known. The progress is possible in two opposite limiting
cases: (i) of a very long, essentially semi-infinite sample, with the sample length $L$ much exceeding the localization length $\xi$ and (ii) of a short metallic sample,
with $L\ll \xi$. We start with analysis of the former case, briefly describing the latter in the end of this section.

\subsubsection{Semi-infinite quasi 1D sample}

Our analysis in this case will be based on the results obtained originally in the work \cite{SO07}, see also \cite{IOS09}. There the so called ``zero mode" solution $ \Psi_0(\lambda,\lambda_1)$ satisfying  $\mathcal{H}\Psi_0=0$ has been found explicitly:
\be \label{zeromode}
  \Psi_0(\lambda,\lambda_1)
  =
  K_0(p) q I_1(q)+p K_1(p) I_0(q) ,
\ee
with
\be \label{pq_def}
  p = \kappa \sqrt{(\lambda_1 +1)/2} ,
\qquad
  q = \kappa \sqrt{(\lambda +1)/2} ,
\ee
where we defined
\be
\label{kappa-def}
  \kappa = \sqrt{8\Gamma/\Delta_\xi} .
\ee
It is clear that such stationary solution corresponds to $t\to \infty$ and  dominates, with exponential accuracy, in the limit of long enough sample lengths:  $L\gg \xi$.
We now proceed with analysing its implications for the resonance statistics in such a limit, relying on Eq.~(\ref{resdenmainM=anyA_text2}).

To this end, it turns out to be convenient to write the resonance density $\rho_E(y)$ as $\rho^{(M)}(y)$, indicating explicitly the number of open channels $M$, and
to define the following generating function:
\be
  \mathcal{Z}(\alpha)
  =
  \sum_{M=1}^\infty
  \rho^{(M)}(y) \alpha^M ,
\ee
which can be easily computed using  Eq.\ (\ref{resdenmainM=anyA_text2}) as
\be \nonumber
  \mathcal{Z}(\alpha)
  =
  \frac{\alpha}{2}
  \int_{-1}^1 d\lambda \, (g-\lambda)
  \sum_{m=0}^\infty
  \frac{[-\alpha(g-\lambda)]^m}{m!}
\frac{\partial^m}{\partial x^m}
\left[ \frac{\frac{\partial^2}{\partial y^2}{\cal F}(\lambda,x;y)}{(x-\lambda)^2}\right]_{x=g}
 \ee
\be \label{genfun}
 =
  \frac{\alpha}{2}
  \int_{-1}^1 d\lambda \, (g-\lambda)
  \frac{\frac{\partial^2}{\partial y^2}{\cal F}(\lambda,x';y)}{(x'-\lambda)^2} = \frac{\alpha}{2(1-\alpha)}
  \int_{-1}^1 d\lambda \,
  \frac{\frac{\partial^2}{\partial y^2}{\cal F}(\lambda,x';y)}{(x'-\lambda)},
\ee
where we used $m=M-1$ and introduced
\be\label{xdash}
  x'=g-\alpha(g-\lambda)=(1-\alpha)g+\alpha\lambda, \quad \mbox{hence} \quad \frac{(g-\lambda)}{(x'-\lambda)}=\frac{1}{1-\alpha}.
\ee

As has been discussed, for the reflection geometry with the lead attached to only one end of a quasi-1D  semi-infinite sample one can write ${\cal F}(\lambda,x;y) = \Psi_0(\lambda,x)$, with identifying $y=2\pi\Gamma/\Delta$. The relation \eqref{kappa-def} implies $y=\frac{\pi \Delta_\xi}{4\Delta}\kappa^2$  hence
\be
  \frac{\partial}{\partial y}{\cal F}(\lambda,x;y)
  =
  \frac{2\Delta }{\pi \Delta_\xi}
  \frac{1}{\kappa}
  \frac{\partial}{\partial\kappa}{\cal F}(\lambda,x;y).
\ee
Now using the standard Bessel functions relations, see e.g. 8.486 in \cite{GRbook}:
\[
\frac{d}{dq}I_0(q)=I_1(q), \, \frac{d}{dp}K_0(p)=-K_1(p), \, \frac{d}{dq}\left[qI_1(q)\right]=qI_0(q), \, \frac{d}{dq}\left[pK_1(p)\right]=-pK_0(p)
\]
as well as $\frac{\partial p}{\partial \kappa}=p/\kappa$ and $\frac{\partial q}{\partial \kappa}=q/\kappa$ we find from \eqref{zeromode}:
\be
\frac{\partial }{\partial \kappa}\Psi_0(\lambda,\lambda_1)=\frac{q^2-p^2}{\kappa}K_0(p) I_0(q)=-\frac{\kappa}{2}(\lambda_1-\lambda)K_0(p) I_0(q)
\ee
giving
 \be
\frac{\partial}{\partial y}{\cal F}(\lambda,x;y)
  =  -
  \frac{\Delta}{\pi\Delta_\xi}
  (x-\lambda) K_0(p) I_0(q) ,
\ee
which when substituted to \eqref{genfun}  implies for the generating function the following explicit representation:
\be
\label{Pi1a}
  \mathcal{Z}(\alpha)
  =
  -
  \left( \frac{2\Delta}{\pi\Delta_\xi} \right)^2
  \frac{1}{\kappa} \frac{\partial}{\partial\kappa}
  \mathcal{O}
, \qquad
  \mathcal{O}
  =
  \frac{\alpha}{1-\alpha}
  \int_0^\kappa
  \frac{1}{\kappa^2}
  K_0(p') I_0(q) q\, dq,
\ee
where we changed in the integral from $\lambda\in[-1,1]$ to $q\in [0,\kappa]$ by $\lambda=2\frac{q^2}{\kappa^2}-1$ and introduced
\be
p'=\kappa \sqrt{\frac{1+x'}{2}}=\sqrt{\alpha q^2+(1-\alpha)p^2}, \quad \mbox{with}\,\,  p=\kappa \sqrt{\frac{1+g}{2}}
\ee
as implied by \eqref{pq_def} and \eqref{xdash}.

Now using the standard integral representation for the Macdonald function $ K_0(p')$, see e.g. 8.432.7 in \cite{GRbook}, we find it is more  convenient to work with
\be\label{La1a}
  \mathcal{O}
  =
  \frac{\alpha}{1-\alpha}
  \int_0^\kappa
  \frac{dq\, q}{\kappa^2} \,
  I_0(q)
  \int_0^\infty \frac{dt}{2t}
  \exp\left[-\frac{t+[p^2-\alpha(p^2-q^2)]/t}{2}\right]
  =
  \sum_{M=1}^\infty \mathcal{O}_M \alpha^M
\ee
Upon further rescaling $t$ by $p$:
\be
\label{La1}
  \mathcal{O}
  =
  \frac{\alpha}{1-\alpha}
  \int_0^\kappa\frac{dq\, q}{\kappa^2}
  I_0(q)
   \,\int_0^\infty \frac{dt}{2t}
  \exp\left[-p\frac{t+1/t}{2}\right]
  \exp\left[\frac{\alpha(p^2-q^2)}{2p t}\right]   \ee
and further using
\be
  \frac{\alpha}{1-\alpha} e^{\alpha\xi}
  =
  \sum_{k=1}^\infty
  \sum_{n=0}^\infty
  \alpha^k
  \frac{\alpha^n\xi^n}{n!}
  =
  \sum_{M=1}^\infty
  \alpha^M
  \sum_{n=0}^{M-1}
  \frac{\xi^n}{n!}
\ee
we can readily extract the required $\mathcal{O}_M$ coefficient defined in \eqref{La1a} from \eqref{La1} as:
\be
  \mathcal{O}_M
  =
  \int_0^\kappa
   \frac{dq\, q}{\kappa^2}I_0(q)
  \int_0^\infty \frac{dt}{2t}
  \sum_{n=0}^{M-1}
  \frac{1}{n!}
  \frac{(p^2-q^2)^n}{(2pt)^n}
  \exp\left[-\kappa\frac{t+1/t}{2}\right].
\ee
 Upon evaluating the integral over $t$ in terms of the Macdonald function one then arrives to a convenient
representation
\be \label{finrho}
\rho^{(M)}(y)=
  -
  \left( \frac{2\Delta}{\pi\Delta_\xi} \right)^2
  \frac{1}{\kappa} \frac{\partial}{\partial\kappa}
  \mathcal{O}_M,
 \qquad  \mathcal{O}_M
  =
  \int_0^\kappa
  \frac{dq\, q}{\kappa^2}
  \sum_{n=0}^{M-1}
  \frac{1}{n!}
  \frac{(p^2-q^2)^n}{(2p)^n}
  K_n(p)
  I_0(q).
\ee
 Recalling the relations $\Delta_{\xi}=D/\xi^2$ and $\xi =2\pi \nu D$ we also see that
\be\label{g_L}
\frac{2\Delta}{\pi\Delta_\xi}=\frac{2}{\pi}\frac{\xi^2}{\nu LD}=\frac{8\pi\nu D}{L}=4g_L
\ee
and in this way arrive to the final explicit expression for the resonance density in quasi-1D semi-infinite sample, valid for any fixed number $M$ of equivalent channels with any channel coupling parameter $g\in[1,\infty]$:
\be
\label{rho-halfwire-gencoup_fin}
  \rho^{(M)}(y)
  =
  -
  \frac{16g_L^2}{\kappa} \frac{\partial}{\partial\kappa}
  \int_0^\kappa
  \frac{dq\, q}{\kappa^2}I_0(q)
  \sum_{n=0}^{M-1}
  \frac{(p^2-q^2)^n}{n!(2p)^n}
  K_n(p)
   ,
\ee
where $p=\kappa \sqrt{(g+1)/2}$. %=\kappa/\sqrt{T}$  and $q = \kappa \sqrt{(\lambda +1)/2}$.
 The above formula is one of central results of our paper.

\subsubsection{ Analysis of Eq.\ (\ref{rho-halfwire-gencoup_fin}) in the regimes of weak and perfect coupling to the leads.}

For long wires in the weak-coupling (tunneling) regime  $g\gg1$ we have $p\gg q$ and Eq.\ (\ref{rho-halfwire-gencoup_fin}) simplifies to
\be
\label{rho-halfwire-tun}
  \rho^{(M)}(y)
  =
  -
  \frac{16g_L^2}{\kappa} \frac{\partial}{\partial\kappa}\left[
  \frac{I_1(\kappa)}{\kappa}
  \sum_{n=0}^{M-1}
  \frac{(p/2)^n}{n!}
  K_n(p)\right],
\ee
where we used $\int_0^{\kappa}dq\,q\,I_0(q)=\kappa I_1(\kappa)$.

In the most important opposite case of the so-called perfect coupling $g=1$ simplifications are possible in view of
the fact that in this case $p=\kappa$. Using the identity 6.657 in \cite{GRbook}:
\be
  \int_0^\kappa
  dq\, q
  (\kappa^2-q^2)^n I_0(q)
  =
  2^n\, n! \kappa^{n+1} I_{n+1}(\kappa)
\ee
we immediately arrive at
\be
\label{L-perf-M}
\rho^{(M)}(y)=
  -
  16g_L^2
  \frac{1}{\kappa} \frac{\partial}{\partial\kappa}\left[
  \frac{1}{\kappa}
  \sum_{n=0}^{M-1}
  K_n(\kappa)
  I_{n+1}(\kappa)\right] .
\ee

\subsubsection{Single-channel limit for perfect coupling.}
In the simplest case of a single-channel lead Eq.\ (\ref{L-perf-M}) takes the form
\be
\label{R-1-res}
  \rho^{(M=1)}(y)= 16g_L^2 R_1(\kappa),
\qquad
  R_1(\kappa)
  =
  -
  \frac1\kappa\frac{d}{d\kappa}
  \frac{K_0(\kappa) I_1(\kappa)}{\kappa}
  =
  \frac{K_1(\kappa)I_1(\kappa) - K_0(\kappa)I_2(\kappa)}{\kappa^2}
  .
\ee

Using the standard asymptotics of modified Bessel functions, see e.g.\ 8.445--8.446 and 8.451 in \cite{GRbook}, one easily finds
\be
\label{R-1-asymp}
  R_1(\kappa)
  =
  \begin{cases}
    (1/2) \kappa^{-2} + (3/8) \ln\kappa + \dots, & \kappa\ll1; \\
    \kappa^{-4} - (3/4)\kappa^{-5} + \dots , & \kappa\gg1.
  \end{cases}
\ee

\begin{figure}
\centering
\includegraphics[width=100mm]{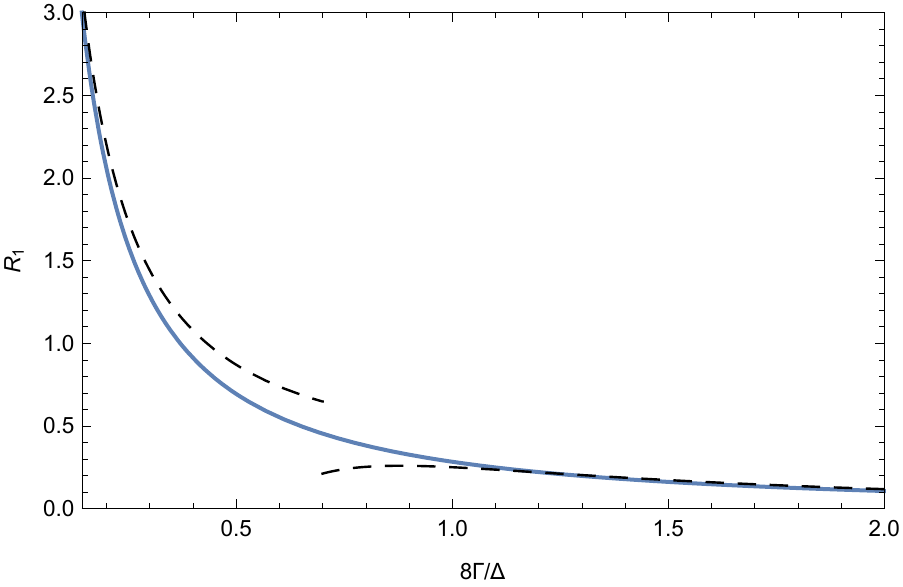}
\caption{$R_1(\kappa)$ for a single-channel reflection from a semi-infinite wire vs.\ $\kappa^2=8\Gamma/\Delta_\xi$ given by \eqref{R-1-res} (solid, blue) and its asymptotic behavior given by \eqref{R-1-asymp} (dashed).}
\label{F:R-wires}
\end{figure}

The energy dependence of the function $R_1(\kappa)$ is shown in Fig.\ \ref{F:R-wires} by the blue (lower) line. In the narrow-widths regime $\kappa\ll1$ we have $R_1\sim1/\kappa^2\sim1/\Gamma$, an anticipated tail behaviour dominated by exponential localization, see the Introduction. The ensuing log-divergence of normalization for $R_1$ stems from the fact that restricting our solution to zero mode amounts to first setting $L\to \infty$, and this limit does not commute with $\Gamma\to 0$.
Indeed, in a long but finite sample of length $L\gg \xi$ the widths of the most narrow resonance is of the order $\Gamma_{min}\sim e^{-L/\xi}$, which sets a cut off of the
logarithmic divergence. In the large-width regime $\kappa\gg 1$ one finds $R_1\sim1/\kappa^4\sim1/\Gamma^2$, which as we shall see is the universal large-width tail behaviour.

\subsubsection{Many-channel limit for perfect coupling: diffusive asymptotics}

In particular, Eq.\ (\ref{L-perf-M}) can be used to describe the explicit intermediate asymptotic behavior of the resonance density in the limit of many perfectly open channels $M=m+1\gg1$ for the regime of resonances with widths large in comparison with the scale $\Delta_{\xi}$, that is $\kappa\gg1$. In such regime it is natural to use the following asymptotic expressions, see e.g. \cite{asybess},
\be
  I_n(x) \approx \frac{e^{S(n,x)}}{\sqrt{2\pi}(n^2+x^2)^{1/4}} ,
\qquad
  K_n(x) \approx \frac{e^{-S(n,x)}}{\sqrt{2/\pi}(n^2+x^2)^{1/4}} ,
\ee
where we defined
\be
  S(n,x) = \sqrt{n^2+x^2} - n\mathop{\rm arcsinh}(n/x)
\ee
 Using this and $m-1\approx M$, one can approximate the sum over $n$ in Eq.\ (\ref{L-perf-M}) by the integral:
\be
  t_M
  \approx
  \frac{1}{\kappa}
  \int_{0}^{M} dn
  \frac{\exp[\partial S/\partial n]}{2\sqrt{n^2+\kappa^2}}\,,
\ee
which upon substituting $\partial S/\partial n = -\mathop{\rm arcsinh}(n/x)$ yields
\be
  t_M
  \approx
  \frac{1}{\kappa}
  \int_{0}^{M} dn
  \frac{1}{2\sqrt{n^2+\kappa^2}}
  \frac{1}{\sqrt{n^2/\kappa^2+1}+n/\kappa}
  =
  \frac{\kappa+M-\sqrt{\kappa ^2+M^2}}{2\kappa^2}
\ee
so that
\be
  -
  \frac{1}{\kappa} \frac{\partial}{\partial\kappa}
  t_M
  \approx
\frac{(\kappa+2 M)\sqrt{\kappa ^2+M^2} -
   (\kappa ^2+2M^2)}{2 \kappa ^4 \sqrt{\kappa ^2+M^2}}
  =
  \begin{cases}
    1/2\kappa^3 , &
     \kappa\ll M
  \\[9pt]
    M/\kappa^4
 , &
     \kappa\gg M
  \end{cases}.
\ee

Thus we see that for $\kappa\gg M$ all terms in the summation give comparable contributions and the result is proportional to $M$ (as can be also seen just by the usual large-argument asymptotics of Bessel functions).
In the opposite limit, $\kappa\ll M$, only $n\sim\kappa$ terms do actually contribute.

Finally we get
\be
\label{rho-res-large}
\displaystyle
  \rho^{(M)}(y)
  \approx
  \left( \frac{2\Delta}{\pi\Delta_\xi} \right)^2
  \frac{(\kappa+2 M)\sqrt{\kappa ^2+M^2} -
   (\kappa ^2+2M^2)}{2 \kappa ^4 \sqrt{\kappa ^2+M^2}}
  =
  \left( \frac{2\Delta}{\pi\Delta_\xi} \right)^2
  \times
  \begin{cases}
    1/2\kappa^3 , &
     \kappa\ll M
  \\[9pt]
    M/\kappa^4
 , &
     \kappa\gg M
  \end{cases}
\ee

\begin{figure}
\centering
\includegraphics[width=100mm]{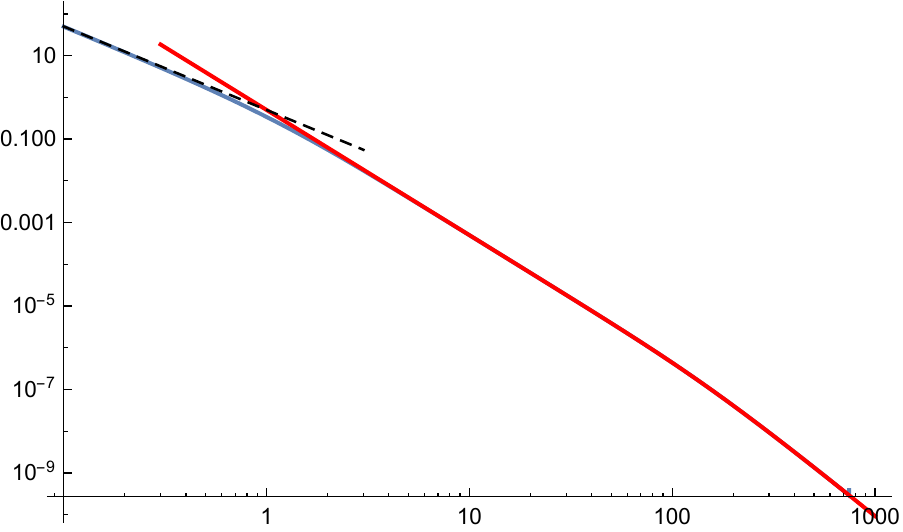}
\caption{Resonance density $\rho^{(M)}(y)$ in units of $(2\Delta/\pi\Delta_\xi)^2$ vs.\ $\kappa$ for $M=100$ perfectly open channels.
Blue: exact expression obtained from \eqref{L-perf-M}.
Red: behavior \eqref{rho-res-large} for $\kappa\gg1$; one can clearly see the intermediate behavior $1/2\kappa^{3}$ for $1\ll\kappa\ll M$. Black dashed: asymptotics $1/2\kappa^2$ in the localized regime.}
\label{F:Pdensity}
\end{figure}

This perfectly fits the exact expression, see Fig.\ \ref{F:Pdensity}. Recall finally that $\kappa\sim \Gamma^{1/2}$, so
the above asymptotics describes the change from the diffusion-dominated intermediate behavior of the density of resonance widths $\rho^{(M)}(\Gamma)\sim \Gamma^{-3/2}$
to the faster universal asymptotic decay $\rho^{(M)}(\Gamma)\sim \Gamma^{-2}$, as was discussed in the Introduction.\\[1ex]

\noindent {\it Summary of the results for a long wire, $L\gg\xi$}. One can easily check that the leading behaviour at $\kappa\ll 1$ for any $M$ is the same as
for $M=1$, and given by the first line in \eqref{R-1-asymp}. Combining everything together, and recalling the relation \eqref{g_L} for $g_L=\xi/L$,
 we see that for a long wire with $g_L\ll 1$ one obtains the following three regimes for
the resonance density:
\be
\label{rho-1D-diff}
  \rho_\text{(1D, long)}(y)
  \approx
  \begin{cases}
    g_L/y ,  & g_Ly \ll 1;
  \\
    g_L^{1/2}/(2y)^{3/2} , & 1 \ll g_Ly\ll M^2;
  \\
    M/4y^2 ,  & M^2 \ll g_Ly.
  \end{cases}
\ee

The appearance of the regime with $\rho(y)\propto y^{-3/2}$ in the case of a large number of channels $M\gg 1$  can be qualitatively understood following the argument of Ref.\ \cite{BGS91}. In a diffusive system, a typical time to leave the sample by reaching the absorbing boundary from a distance $r$ is $t\sim r^2/D$. Associating the level widths with inverse times and using $\rho(\Gamma)\,d\Gamma\sim dr/L$, we immediately get
the density  $\rho(\Gamma) \sim \sqrt{D}/(L\Gamma^{3/2})$, in accordance with our analytic results. Note however that the above reasoning is essentially relying only on
the classical diffusion picture inside the sample, completely disregarding the issue of channel coupling. Hence, if applied naively the same argument would wrongly predict the  $\Gamma^{-3/2}$ asymptotics to hold for any $M$. It also misses completely  the last, \mbox{(quasi-)ergodic} regime of large widths.
Our rigorous analysis therefore makes it clear that only if {\it both} the propagation inside the media {\it and} the decay into channels at the boundary can be treated semi-classically on equal footing (which is essentially ensured only by the condition $M\gg 1$)
the resonance statistics can be explained by appealing to semiclassical arguments.

\subsubsection{Metallic quasi 1D sample: transfer matrix solution}
\label{S:3.2.5}
So far we considered the localization regime $L\gg \xi$. Finally, we may wish to consider the opposite limiting case: a finite piece of quasi-1D metal of length $L\ll\xi$. We further assume the parameter regime $\Gamma\gg\Delta$. In this (perturbative) regime it is well-known that the sigma-model manifold can be treated as essentially ``flat", and the function $\Psi(\lambda,\lambda_1)$ will be dominated by a small vicinity of the origin in Cartan coordinates: $\lambda_1=\lambda=1$.
In this situation it is more convenient to rewrite the transfer-matrix Hamiltonian \eqref{Ham-lambda} in the angular coordinates $\lambda_1=\cosh \theta_1$ and $\lambda=\cos \theta_2$, and then expand in small $\theta_1\ll1 $ and $\theta_2\ll 1$.  In this way one gets:
\be
\label{Ham-theta}
  \mathcal{H}
  =
  -
  %\frac{1}{2}
  \left(
    \frac{\partial^2}{\partial \theta_1^2}
  +
    \frac{\partial^2}{\partial \theta_2^2}
  \right)
  -
  %\frac{1}{2}
  \left(
    \frac{1}{\theta_1}
    \frac{\partial}{\partial \theta_1}
  +
    \frac{1}{\theta_2}
    \frac{\partial}{\partial \theta_2}
  \right)
  +
  \frac{4}{\theta_1^2+\theta_2^2}
  \left(
    \theta_1 \frac{\partial}{\partial \theta_1}
  +
    \theta_2 \frac{\partial}{\partial \theta_2}
  \right)
  +
  \frac{\Gamma}{2\Delta_\xi} (\theta_1^2+\theta_2^2) .
\ee

Now we seek for the solution of the transfer-matrix  Schr\"odinger-like equation \eqref{transf-eq} in the form
\be
\label{Psi(xyt)}
  \Psi(\theta_1,\theta_2;t) = \exp[-C(t)(\theta_1^2+\theta_2^2)/2] .
\ee
Using the flat Hamiltonian \eqref{Ham-theta} we see that $H\Psi = [ \Gamma/2\Delta_\xi - C^2 ] (\theta_1^2+\theta_2^2) \Psi$ and therefore this Ansatz satisfies the equation provided the function $C(t)$ obeys
\be
  \frac12 \frac{\partial C}{\partial t}
  =
  \Gamma/2\Delta_\xi - C^2
  =
  \kappa^2/16 - C^2 ,
\ee
where $\kappa$ is defined in \eqref{kappa-def}.

Solving with the initial condition $\Psi=1$ at $t=0$ we find
\be
\label{alpha-res}
  C(t)
  =
  \frac{\kappa}{4}
  \tanh \frac{\kappa t}{2}
  =
  \sqrt{\frac{\Gamma}{2\Delta_\xi}}
  \tanh   \sqrt{\frac{2\Gamma}{E_\text{Th}}} ,
\ee
where $E_\text{Th} = D/L^2$ is the Thouless energy.
An equivalent expression can be found in Eq.\ (24) of \cite{Micklitz}, where it was derived using a mapping to the Coulomb problem \cite{SO07}, which looks to us as a somewhat less straightforward derivation than given above. Yet another derivation will be given in the framework of perturbative approximation to the nonlinear sigma-model described in the next section, where
implications of such form for resonance density in quasi 1D metallic samples will be discussed as well.

\subsection{Good metal limit: perturbative treatment in any dimension}

In a good metal, the OPF can be calculated via the perturbative treatment
of the nonlinear $\sigma$-model which we sketch below following the book \cite{Efe_book}. In such a regime fluctuations of supermatrices $Q$ around
the supermatrix $\Lambda=\diag(1,1,-1,-1)\equiv \diag(\mathbf{1},-\mathbf{1})$ are expected to be small, and to exploit this smallness
it is convenient first to parametrize the sigma-model manifold around the point $\Lambda$ as $Q=\Lambda (\mathbf{1}+iP)(\mathbf{1}-iP)^{-1}$ in terms of supermatrices $P$ of the form $P=\begin{pmatrix} 0 & B\\ \overline{B} & 0\end{pmatrix}$. After that one expands in $P$, retaining in the leading approximation only the two lowest order terms:
\be
\label{1pert}
Q=\Lambda (\mathbf{1}+iP)(\mathbf{1}-iP)^{-1}\approx \Lambda (\mathbf{1}+2iP-2P^2+\ldots)
\ee
In this approximation the continuum sigma-model action, Eq.\ (\ref{nlsigmacont}), takes the form
\begin{equation}\label{nlsigma_Gau}
{\cal S}[P]
\approx
-\pi\nu\int d^d r \Str\left[D(\nabla P)^2 +2 \eta\, P^2\right],
\end{equation}
whereas the measure of integration is essentially ``flat", i.e.\ $d\mu(Q)=dP$, making the corresponding theory purely Gaussian.
The integral defining the OPF via Eq.\ \eqref{OPF-def} is then straightforwardly evaluated, yielding
\begin{equation}\label{OPF-def_pert}
F_{\bf r}(Q;\eta)=\int_{P_{\bf r}=P}\prod_{{\bf r}'\ne {\bf r}} dP_{\bf r}'\, \exp(-{\cal S}[P])=
\exp\left(c(\mathbf{r}; \eta)\Str P^2\right) ,
\end{equation}
where we introduced
\be
 \quad c(\mathbf{r};\eta) = 1/\Pi(\mathbf{r}, \mathbf{r}, \eta),
\ee
with
\be
\Pi(\mathbf{r}, \mathbf{r}, \eta) = \corr{\mathbf{r}|(-D\nabla^2+2\eta)^{-1}|\mathbf{r}}/\pi\nu
\ee
being the so-called diffusion propagator with Neumann boundary conditions, corresponding to an isolated sample.
For a hypercube of size $L$ with the waveguide coupled to its corner one then has the following representation:
\begin{equation}\label{difpropcorner}
\Pi(\mathbf{0},\mathbf{0},\eta)=\frac{\Delta}{\pi}\sum \frac{1}{D\mathbf{q}^2+2\eta} ,
\end{equation}
with the summation running over momenta $\mathbf{q}=(\pi/L)(n_1,n_2,...n_d)$, with non-negative integer $n_i$. A detailed analysis of
$\Pi(\mathbf{0},\mathbf{0},\eta)$ is provided in \ref{Appendix}.

 Further using (\ref{1pert}) to eventually restore $Q$ in \eqref{OPF-def_pert} one finally arrives at the OPF in the form
\begin{equation}\label{OPF-def_pertA}
F_{\bf r}(Q;\eta)\approx \exp\left[-\frac{1}{2}c(\mathbf{r}; \eta)\Str \left(Q_{\mathbf{r}}\Lambda\right)\right],
\end{equation}
 Denoting for brevity $c(\eta):=c(\mathbf{r};\eta)$ we therefore get the OPF in the standard
parametrization of $Q$ matrices in terms of the compact and noncompact Cartan coordinates $(\lambda,\lambda_1)$ as
${\cal F}(\lambda,\lambda_1;\eta)=e^{-c(\eta)(\lambda_1-\lambda)}$. The validity of this perturbative expression for the OPF is
restricted by the condition $c(\eta)\gg 1$. Notice of a close similarity of the perturbative OPF to the previously considered 0D case.

Identifying $\eta\to \Gamma$ and substituting the perturbative OPF into the main formula Eq.~(\ref{resdenmainM}) we can analyse the ensuing mean density
of properly scaled resonance widths $y=2\pi \Gamma/\Delta$ in terms of the inverse diffusion propagator $c(y)$. We will consider separately the single open channel case $M=1$ and the case of many channels $M\gg 1$.

\subsubsection{Single open channel $M=1$}
The perturbative expression for the resonance density follows from substituting the found OPF to Eqs.\ (\ref{resdenmainM}) and can be evaluated as
\begin{multline}
\label{den_pert_1a}
\rho^{(M=1)}(y)\approx \frac{1}{2}\int_{-1}^1(g-\lambda)\frac{\partial^2}{\partial y^2}\frac{e^{-c(y)(g-\lambda)}}{(g-\lambda)^2}d\lambda
\\{}
=
-\frac{1}{2}\frac{\partial}{\partial y}\left[\frac{c'(y)}{c(y)}\left(e^{-(g-1)c(y)}-e^{-(g+1)c(y)}\right)\right]
\approx -\frac{1}{2}\frac{\partial}{\partial y}\left[\frac{c'(y)}{c(y)}\,e^{-(g-1)c(y)}\right] ,
\end{multline}
where we exploited $c(y)\gg 1$. Especially simple expression arises for the perfect coupling case:
\be\label{den_pert_1c-1}
\rho^{(M=1)}(y) \approx -\frac{1}{2}\frac{\partial^2}{\partial y^2}\ln c(y)= \frac{1}{2}\frac{\partial^2}{\partial y^2}\ln\Pi(0,0,y),
\ee
where we assumed for simplicity that the waveguide is coupled to the corner of disordered sample, which we choose as the origin.

\subsubsection{Many open channels $M\gg 1$}
In this case it is convenient first to rewrite  the main formula Eqs.\ \eqref{resdenmainM} in an equivalent form
via a contour integral over complex variable $\zeta$ replacing the non-compact real variable $\lambda_1$:
\begin{equation}\label{resdenmainM=anyA_contour}
  \rho^{(M)}(y)
  =
  \frac{(-1)^{M-1}}{2}\frac{\partial^2}{\partial y^2}
  \int_{-1}^1(g-\lambda)^M\,
  \,d\lambda \oint_{\mathcal{C}}\, \frac{{\cal F}(\lambda,\zeta;y\Delta/2\pi)}{(\zeta-g)^M(\zeta-\lambda)^2}\frac {d\zeta}{2\pi i},
\end{equation}
where the contour ${\mathcal{C}}$ encircles the point $\zeta=g$ on the real axis. Substituting here the perturbative OPF ${\cal F}(\lambda,\zeta;\eta)=e^{-c(y)(\zeta-\lambda)}$ one can further bring (\ref{resdenmainM=anyA_contour}) to the form
\begin{equation}\label{resdenmainM=anyA_contour_pert1}
  \rho^{(M)}(y)
  =
  \frac{(-1)^{M}}{2}\frac{\partial}{\partial y}\left[ c'(y) P_M(y)\right], \quad
  P_M(y)=\int_{-1}^1(g-\lambda)^M\,
  \,d\lambda \oint_{{\mathcal{C}}}\, \frac{e^{-c(y)(\zeta-\lambda)}}{(\zeta-g)^M(\zeta-\lambda)}\,\frac{d\zeta}{2\pi i} ,
\end{equation}
 The latter relation immediately implies
 \be
 \frac{\partial }{\partial y}P_M(y)=-c'(y)\,J_1(c(y))\,J_2(c(y)), \quad J_1(c)=\int_{-1}^1(g-\lambda)^M\,e^{c\lambda}\,d\lambda, \quad
  J_2(c)=\oint_{{\mathcal{C}}} \frac{e^{-c\zeta}}{(g-\zeta)^M}\,\frac{d\zeta}{2\pi i}
 \ee
 In the limit $M\gg1$ the integrals $J_1(c)$ and $J_2(c)$ are dominated by the saddle-points at $\lambda=\xi=g-\frac{M}{c}$
 and simple calculation yields the corresponding leading order the contribution:
 \be
 J_1(c)\,J_2(c)\approx (-1)^M\frac{ M}{c^2}, \quad \mbox{if}\quad \frac{M}{g+1}<c< \frac{M}{g-1}
 \ee
 and exponentially small otherwise. As  $c(y)\to \infty$ for $y\to \infty$, the above allows to  restore $P_{M\gg 1}(y)$ in the form
 \be
 P_{M\gg 1}(y) \approx -  (-1)^M M\int_y^{\infty}\frac{c'(\tilde{y})}{c^2(\tilde{y})}\,d\tilde{y}= (-1)^M \frac{M}{c(y)}
 \ee
 implying for the resonance density for $M\gg 1$ the following expression:
 \be \label{den_pert_1c}
\rho^{(M)}(y) \approx -\frac{M}{2}\frac{\partial^2}{\partial y^2}\ln c(y) =\frac{M}{2}\frac{\partial^2}{\partial y^2}\ln \Pi(0,0,y), \quad  \frac{M}{g+1}<\frac{1}{\Pi(0,0,y)}< \frac{M}{g-1}
\ee
 We thus see that in the  good metal regime the resonance density in the case of perfect coupling $g=1$ for $M=1$  and $M\gg 1$ is described essentially by the same expression in terms of  the diffusion propagator, up to an overall factor $M$. We therefore need to understand  the functional dependence $c(y)$ in different spatial dimensions, which amounts to studying the corresponding diffusion propagators. This is done in \ref{Appendix} to the paper, and below we very briefly discuss the ensuing resonance densities.    \\[0.5ex]

\noindent {\it Perturbative regime in quasi one-dimensional samples.} Consider first the case of a quasi-1D disordered system, physically describing a diffusive
wire of length $L$ supporting many propagating transverse modes. The relevant  parameters are the wire conductance $g_L=\xi/L$, the localization length $\xi=2\pi\nu D$,  the mean level spacing $\Delta=1/{\nu L}$ and the Thouless energy $E_\text{Th}=D/L^2$. The perturbative (Gaussian) approximation is valid provided $y\gg \mbox{max}\{1, g_L^{-1}\}$.
 One finds in this case $c(\eta)=(g_L/2) f(2\eta/E_\text{Th})$ with $f(z)=\sqrt{z}\tanh\sqrt{z}$. An alternative derivation of this result, using transfer-matrix language specific for the quasi-1D geometry, has been given by us earlier in the framework of transfer-matrix formalism, see Section \ref{S:3.2.5}.

 Assuming perfect coupling condition $g=1$, for a short metallic wire ($L\ll\xi$ or, equivalently, $g_L\gg 1$), the resonance density
$\rho_\text{(1D,short)}(y)$ can be found from \eqref{den_pert_1c}. Replacing $\eta\to \Gamma=\Delta y/2\pi$ one gets for $c(y)$ the expression
  $c(y)=(g_L/2) f(2y/g_L)$, implying  $c(y)\approx y$ for $y\ll g_L$ and $c(y)\approx \sqrt{yg_L/2}$ for $y\gg g_L$. We then see
that the resonance density in this case interpolates between  $\rho_\text{(1D,short)}(y)\approx M/2y^2$ (full ergodization, 0D case, RMT result)  for $1\ll y\ll g_L$
and  $\rho_\text{(1D,short)}(y)\approx M/4y^2$ (partial ergodization) at $y\gg g_L$. Both regimes of $\rho(y)$ are pretty universal, with the diffusion coefficient dropping from the answer.\\[0.5ex]
\noindent {\it Perturbative regime in a good metal in $d=3$.}  Following the same lines as above, we find using results from \ref{Appendix} in the case of 3-dimensional situation:
\be
\label{rho-3D-diff_text}
  \rho_\text{(3D)}(y)
  \approx
  \frac{M}{2}\left[\frac{1}{y^2}-\frac{1}{(y+\pi g_l)^2}\right],
\ee
where $g_l=2\pi\nu D l\gg 1$ is the system's microscopic conductance at the scale of the mean free path $l$. Equation (\ref{rho-3D-diff_text}) interpolates between the RMT limit at $1\ll y\ll g_l$ and $  \rho_\text{(3D)}(y)=g_l\pi/y^3$ at $g_l\ll y\ll g_L$. Similar consideration can be done in $d=2$ as well, see \ref{Appendix}.

%%%%%%%%%%%%%%%%%%%%%%%%%%%%%%%%%%%%%%%%%%%%%%%%%%%%%%%%%%%%%%%%%%%%%%%%%%%%%%%%%%%%%%%%%%
\section{Conclusions, Summary and Discussion}
\label{S:4}

\subsection{Main results}

In this paper, starting with the problem of wave reflection from a disordered medium via a single
$M$-channel lead, we have revealed a general relation between the mean density of $S$-matrix poles (resonances) and the order parameter function characterizing the medium in the nonlinear sigma-model approximation. This relation allowed us to find the mean resonance density explicitly in various regimes, including the non-perturbative localization regime in a quasi-one-dimensional geometry. In particular, for $M\gg 1$ we provided a fully controlled {\it ab initio} derivation of the diffusive intermediate asymptotics $\rho(\Gamma)\sim 1/\Gamma^{3/2}$ anticipated by qualitative arguments \cite{BGS91}. Note that some experiments related to studies of zeroes and poles of the scattering matrix in quasi-1D disordered samples have been recently reported \cite{Genack_etal_2022}.

Below we give a summary of the main results in different physically interesting limits, and then report the results of numerical simulations in some of these regimes, and beyond.
\begin{enumerate}
\item {\it Zero-dimensional (0D) limit.} The simplest case is that of
a fully ergodic RMT-like system, where we recover the resonance density $\rho_\text{(0D)}(y),\,\,y=2\pi \Gamma/\Delta$ obtained earlier by several other method in Refs.\ \cite{{Fyo96}} and \cite{FyoKhor99}.
%\be
%\label{rho-0D-gen}
%  \rho_\text{(0D)}(y)
%  =
%  \frac{(-1)^{M}y^{M-1}}{(M-1)!}\frac{\partial^M}{\partial y^M}\left(e^{-yg}\frac{\sinh{y}}{y}\right) ,
%\ee
In the perfect coupling limit, the resonance density has a maximum at $y\approx M/2$, with a tail $\rho_\text{(0D)}(y) \approx M/2y^2$ at $y>M/2$.

\item {\it Quasi-1D wires}. Next case is a diffusive
wire of length $L$ supporting many propagating transverse modes.
The relevant  parameters are the wire conductance $g_L=\xi/L$, with the localization length $\xi=2\pi\nu D$ related to the classical diffusion coefficient $D$.
%The mean level spacing $\Delta=1/{\nu L}$ and the Thouless energy $E_\text{Th}=D/L^2$.
The $M$-channel lead is attached to one end of the wire, with the parameter $1\le g<\infty$ characterizing
the coupling strength between the lead and the wire, and considered to be the same across all channels, with minimal $g=1$ corresponding to the
 perfect-coupling case.

 For a {\it short wire} ($L\ll\xi$ or equivalently $g_L\gg 1$), the resonance density
$\rho_\text{(1D,short)}(y)$ can be found by perturbative methods in the diffusion approximation. It interpolates between $\rho_\text{(1D),short}(y)\approx M/2y^2$ (full ergodization, 0D case, RMT result) at $1\ll y\ll g_L$ and $\rho_\text{(1D,short)}(y)\approx  M/4y^2$ (partial, or quasi-ergodization) at  $y\gg g_L$. Such behaviour looks quite universal.

The diffusive approximation breaks down for long wires ($L\gg\xi$, or $g_L\ll 1$) at $y<1/g_L$ that requires a nonperturbative solution  of the nonlinear sigma model via a transfer matrix technique, which we implement following the work \cite{SO07}. In this regime it is convenient to introduce a dimensionless parameter $\kappa$ related to the resonance width $\Gamma$ as
\be
  \kappa^2 = 8\Gamma/\Delta_\xi = 8g_L y,
\ee
with $\Delta_\xi = D/\xi^2$ being the characteristic level spacing in the sample whose length $L$ coincides with the localization length $\xi$.
The value of $\kappa$ distinguishes between the localized ($\kappa\ll1$) and diffusive ($\kappa\gg1$) regimes.
 The following explicit expression for the resonance density is then obtained  in terms of the modified Bessel functions for any channel coupling parameter $g\in[1,\infty]$:
\be
\label{rho-halfwire-gencoup}
  \rho_\text{(1D)}(y)
  =
  -
  \frac{16g_L^2}{\kappa} \frac{\partial}{\partial\kappa}
  \int_0^\kappa
  \frac{dq\, q}{\kappa^2}
  \sum_{n=0}^{M-1}
  \frac{(p^2-q^2)^n}{n!(2p)^n}
  K_n(p)
  I_0(q) ,
\ee
where $p=\kappa \sqrt{(g+1)/2}$.

In the special case of perfect coupling $g=1$ we have $p=\kappa$ and \eqref{rho-halfwire-gencoup} simplifies to
\be
\label{rho-wire-perfect-M}
  \rho_\text{(1D)}(y)
  =
  -
  \frac{16g_L^2}{\kappa} \frac{\partial}{\partial\kappa}
  \frac{1}{\kappa}
  \sum_{n=0}^{M-1}
  K_n(\kappa)
  I_{n+1}(\kappa) .
\ee
Asymptotic behavior of this analytic expression is summarized below:
\be
\label{rho-1D-diff_concl}
  \rho_\text{(1D,long)}(y)
  \approx
  \begin{cases}
    g_L/y ,  & y \ll 1/g_L;
  \\
    g_L^{1/2}/(2y)^{3/2} , & 1/g_L \ll y\ll M^2/g_L;
  \\
    M/4y^2 ,  & M^2/g_L \ll y.
  \end{cases}
\ee
It reproduces the perturbative (diffusion approximation) results in the second and third lines, and provides a nonpertubative answer in the limit of small widths ($\Gamma\ll\Delta_\xi$, the first line). In the latter regime, the resonance density $\rho(y)$ scales as $1/y$, indicating the presence of states very weakly coupled to the lead. These are the states localized far away from the wire edge and decaying exponentially away, which are mainly responsible for the normalization of $\rho(y)$ (logarithmic divergency of the normalization integral at small $y$ is cured by the absence of widths smaller than $\Gamma_\text{min}\sim\Delta_\xi e^{-2L/\xi}$, corresponding to the state localized at the edge opposite to the attached lead).

The appearance of the regime with $\rho(y)\propto y^{-3/2}$ in the case of a large number of channels $M\gg 1$ [the second line of Eq.\ (\ref{rho-1D-diff_concl})] can be qualitatively understood following the argument of Ref.\ \cite{BGS91}. One may actually see that the crossover between the second (diffusive) and the last (quasi-ergodic) regime in \eqref{rho-1D-diff} happens at $y_\textrm{erg}\sim M^2/g_L$, which together with the relation $g_L\sim \Delta \xi^2/D$ implies the resonance widths estimate $$\Gamma_{\text{erg}}\sim y_\textrm{erg}\Delta \sim M^2 D/\xi^2 \sim  D/L_\text{erg}^2,$$
and introduces a new spatial scale $L_\text{erg}=\xi/M\ll \xi$. The quasi-ergodic regime of decay then corresponds to the states localized close enough to
the lead, at distances $L < L_\text{erg}$, whereas those states localized in the interval of length $L_\text{erg} < L\ll \xi$ decay diffusively,
and for those even further away at $L\gg \xi$ localization effects play the major role and are responsible for the first regime in \eqref{rho-1D-diff_concl}.

\item {\it A good 3D metal}.  Neglecting the  localization effects, one again can use the diffusion approximation,  and find
\be
\label{rho-3D-diff}
  \rho_\text{(3D)}(y)
  \approx
  \frac{M}{2}\left[\frac{1}{y^2}-\frac{1}{(y+\pi g_l)^2}\right],
\ee
where $g_l=2\pi\nu D l\gg 1$ is the system's microscopic conductance at the scale of the mean free path $l$. Equation (\ref{rho-3D-diff}) interpolates between the RMT limit at $1\ll y\ll g_l$ and $  \rho_\text{(3D)}(y)=g_l\pi/y^3$ at $g_l\ll y\ll g_L$.

\end{enumerate}

\subsection{Numerical study}

\begin{figure}
\centering
\includegraphics[width=0.80\textwidth, angle=00]{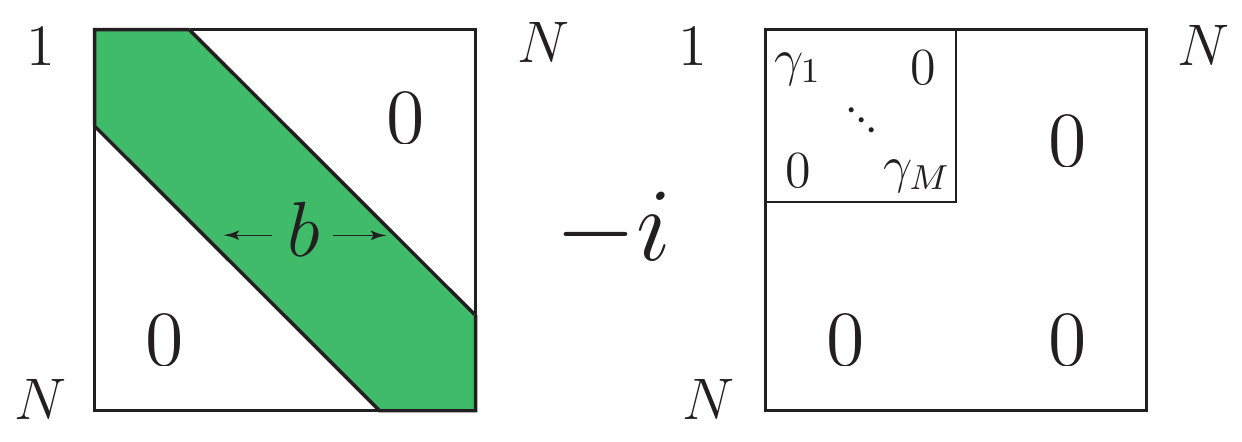}
\caption{Schematic structure of the random matrix model, Eq.\ \eqref{heff}.}
\label{BRM}
\end{figure}

As has been already mentioned, a convenient way which allows to test numerically analytical predictions obtained in the framework of Efetov nonlinear $\sigma$-model amounts to replacing the Hamiltonian $H$ for the disordered part with an appropriately chosen Hermitian Banded Random Matrix (BRM), corresponding to  broken time reversal symmetry.
To this end we will compute eigenvalues of the effective non-Hermitian Hamiltonian
\begin{equation}
\label{heff}
{\cal H}_\text{eff}=H-(i/\pi\nu)\,\mbox{diag}(\mathbf{1}_M,0_{N-M})
\end{equation}
corresponding to the case of equivalent perfectly coupled channels: $g_c=1$ for all $c=1,\ldots, M$
and  work in the middle of the spectrum considering $10\%$  of the eigenvalues closest to $0$.
For dimension $d=1$ the structure of the associated non-Hermitian random matrix is sketched in the Fig.\ \ref{BRM}.
Similar constructions can be used also for  $d=2,3$. The ensemble of BRM in $d$ dimensions is defined over a hyper-cubic lattice of linear size $L$ and is specified by the bandwidth $b$. The entries are normally distributed complex numbers $H_{\mathbf{r}_1\mathbf{r}_2}=H_{\mathbf{r}_2\mathbf{r}_1}^*$, where ${\bf r}_{1,2}\in \mathbb{Z}^d$, with zero mean and the variance % (here $|{\cdot}|$ is the $L_1$ norm)
\begin{equation}
%%\left<H_{\mathbf{r}_1\mathbf{r}_2}\right>=0,\quad
\left<|H_{\mathbf{r}_1\mathbf{r}_2}|^2\right>
=
\Theta(b-|\mathbf{r}_1-\mathbf{r}_2|)
(1+\delta_{\mathbf{r}_1\mathbf{r}_2})/2.
\end{equation}

\begin{figure}[b!]
\centering
\includegraphics[width=100mm]{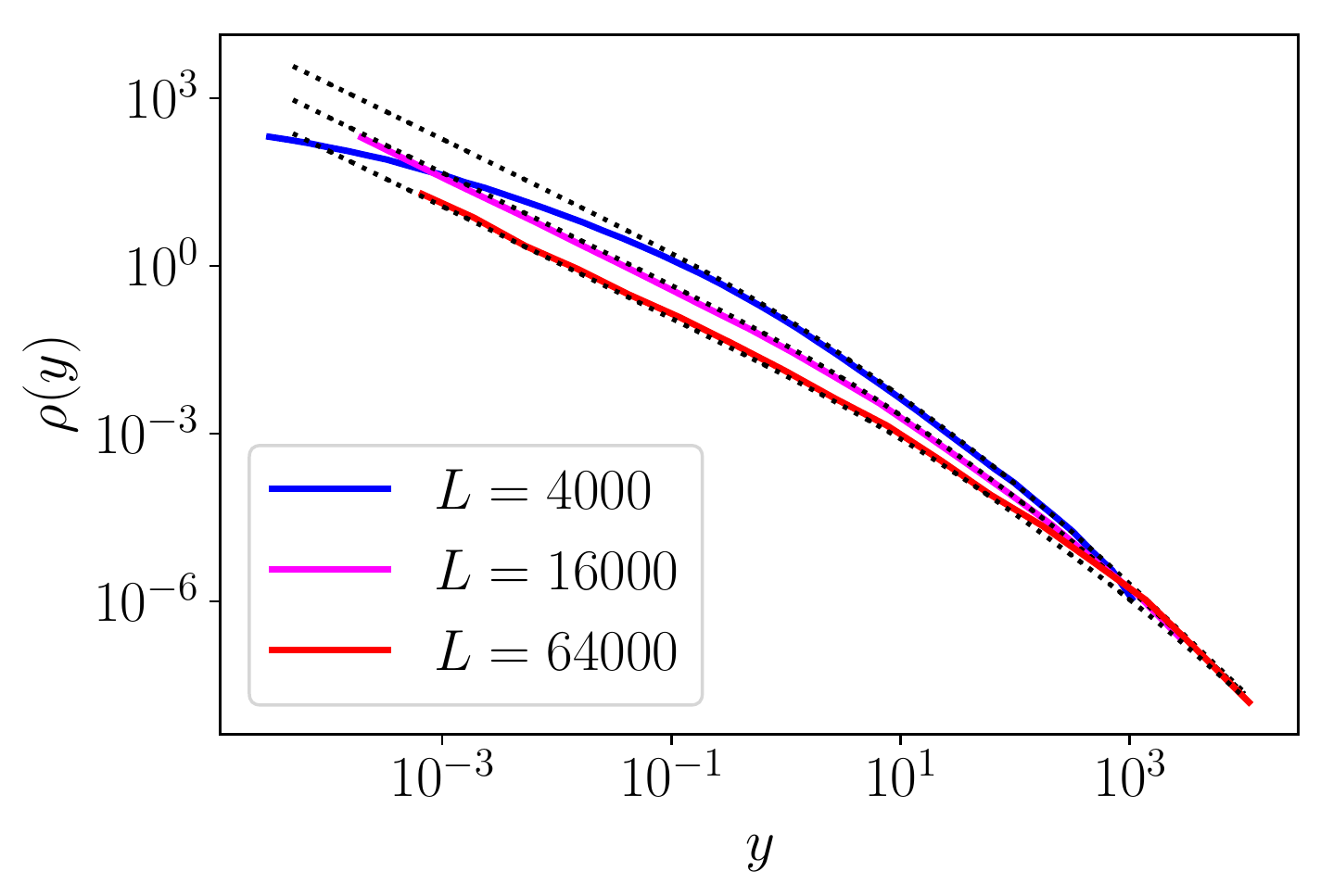}
\caption{Density of resonances $\rho_\text{1D}(y)$ computed for BRM ($b=30$, corresponding to $\xi=600$) with $M=10$ channels, perfectly coupled to the edge. Solid: exact diagonalization for various system sizes (indicated on the legend). Dashed: exact analytical result given by Eq.\ \eqref{rho-wire-perfect-M}.}
\label{F:Q1D}
\end{figure}

For $d=1$, this model belongs to the universality class of quasi-1D localization \cite{FyoMirRBM91,FyoMirRBM94,SchSch19,SchSch21}, with the localization length (in the middle of the spectral band) $\xi = 2b^2/3$~\cite{fyodorov1993}. For the numerical simulation, we pick the bandwidth $b=30$ and attach a lead with $M=10$ channels to the end-point of the wire. By exact diagonalization of the effective Hamiltonian \eqref{heff} we evaluate the resonance density $\rho_\text{(1D)}(y)$ for several values of $L\gg\xi$, see Fig.\ \ref{F:Q1D}. For the largest system sizes our numerical results perfectly follow Eq.\ (\ref{rho-wire-perfect-M}) applicable in the limit $L\to\infty$ (dashed lines), while for the smallest wire length, yet $L\gg\xi$, the density $\rho_\text{(1D)}(y)$ clearly deviates from Eq.\ \eqref{rho-wire-perfect-M}, with an apparent saturation at $y\lesssim \left(L/\xi\right) e^{-2L/\xi} \sim 10^{-5}$.

\begin{figure}
\centering
\includegraphics[width=100mm]{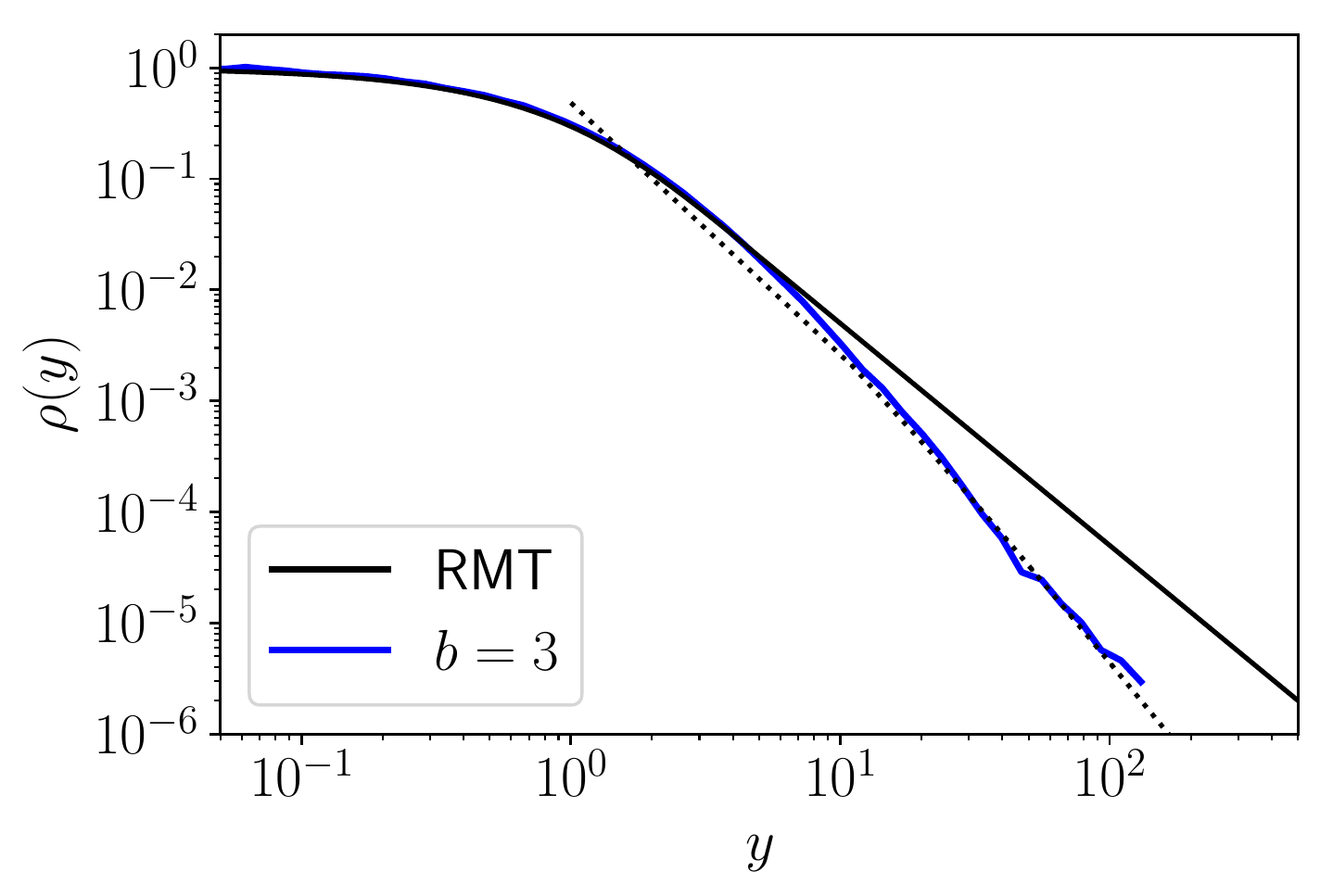}
\caption{Density of resonances $\rho_\text{3D}(y)$ computed numerically for $d=3$ BRM model of a linear size $L=30$ and $M=1$ channel attached to the corner of the lattice, with the bandwidth $b=3$. The RMT result is shown by the solid black line, the dashed line corresponds to Eq.\ (\ref{rho-3D-diff}) with $g_l=1.5$.}
\label{F:Q3D}
\end{figure}

For $d=3$, the result of simulation of the BRM model at $L=30$ and bandwith
$b=3$ is shown on the Fig. \ref{F:Q3D}. We consider a single-channel lead ($M=1$) which is coupled to a sphere of radius $r=3$ (we have found that coupling to a single node is not sufficient to ensure applicability of the $\sigma$-model description). In the same figure we show with dashed lines the fit according to Eq.\ (\ref{rho-3D-diff}), where conductance is considered as a fitting parameter, leading to $g_l=1.5$.

Considering perspectives for further research, extensions of the sigma-model treatment of resonance statistics to disordered media with preserved time reversal invariance should be certainly possible. Further, it would be very interesting to understand analytically critical features of the resonance density in the vicinity of the Anderson localization transition \cite{Kott02}.  From such a perspective studying the Anderson model on the Random Regular Graph  \cite{Tikh_Mir_Skvor_2016} where some features of the associated OPF have been analytically addressed in recent years \cite{Tikh_Mir_2019} represents a promising direction. In addition, one may hope that the recent numerical results \cite{scattPRBM} about critical resonance scattering in the framework of the model of Powerlaw BRM should be amenable to analytic investigation in the present framework.
 Another interesting avenue is to study resonances in disordered media beyond nonlinear sigma-model framework, e.g., for finite number of transverse channels  using the  DMPK-type approach to scattering with absorption \cite{Bruce_Chalker_absorp_DMPK_1996,Misir_Paas_Beenak_absorp_DMPK_1997}. Some results in this direction will be published elsewhere \cite{Fyo_Meibohm_progr}. It would be certainly also interesting to attempt to understand other related questions, like statistics of residues at $S$-matrix poles, reflecting the effects of eigenfunction non-orthogonality in open disordered systems, as well as to characterize correlations between positions of neighboring poles in the complex plane in different regimes.

\section*{Acknowledgements}

Y.V.F. acknowledges financial support from EPSRC Grant EP/V002473/1 ``Random Hessians and Jacobians: theory and applications''.

%%%%%%%%%%%%%%%%%%%%%%%%%%%%%%%%%%%%%%%%%%%%%%%%%%%%%%%%%%%%%%%%%%%%%%%%%%%%%%%%%%%%%%%%%%
%%%%%%%%%%%%%%%%%%%%%%%%%%%%%%%%%%%%%%%%%%%%%%%%%%%%%%%%%%%%%%%%%%%%%%%%%%%%%%%%%%%%%%%%%%
%%%%%%%%%%%%%%%%%%%%%%%%%%%%%%%%%%%%%%%%%%%%%%%%%%%%%%%%%%%%%%%%%%%%%%%%%%%%%%%%%%%%%%%%%%

\vspace{1cm}

\appendix

\section{Diffusion propagators for different dimensions}
\label{Appendix}

In our explicit calculations we for simplicity consider an isolated sample of random medium in the form of a hypercube of size $L$ with the waveguide coupled to its corner. The diffusion propagator, proportional to the Green's function of the diffusion equation, can be found in a given spatial dimension e.g. by finding all eigenfunctions of the classical diffusion equation  with Neumann boundary conditions.

\subsection{$d=1$}

First we consider the simplest case $d=1$ where the complete orthonormal set of eigenfunctions $y_n(x), x\in[0,L]$ solving $\left(-D\frac{d^2}{dx^2}+2\eta\right)y_n(x)=\lambda_n y_n(x)$ with the Neumann conditions $y_n'(0)=y'_n(L)=0$ is given by
\be \label{Neumann_eigen}
y_n(x)=\left\{\begin{array}{cc}\sqrt{\frac{2}{L}}\cos{\left(\frac{\pi n x}{L}\right)}, & n=1,2,\ldots \\
\sqrt{\frac{1}{L}}, & n=0\end{array}\right., \qquad \lambda_n= D \frac{\pi^2 n^2}{L^2}+2\eta.
\ee
The corresponding Green's function solving  $\left(-D\frac{d^2}{dx^2}+2\eta\right)G(x,x';\eta)=\delta(x-x')$
is then readily given by
\be \label{Neumann_Green}
G(x,x';\eta)=\frac{1}{2\eta L}+\frac{2}{L}\sum_{n=1}^{\infty}
\frac{\cos{\left(\frac{\pi n x}{L}\right)} \cos{\left(\frac{\pi n x'}{L}\right)}}{D \frac{\pi^2 n^2}{L^2}+2\eta}
\ee
which for the required diffusion propagator immediately gives:
\be \label{propag_1d}
\Pi(0,0;\eta)=\frac{1}{\pi \nu L}\left[\frac{1}{2\eta }+2\sum_{n=1}^{\infty}
\frac{1}{D \frac{\pi^2 n^2}{L^2}+2\eta}\right]=\frac{1}{\pi \nu}\,\sqrt{\frac{1}{2\eta D}}\,
\coth{\left(L\sqrt{2\eta/D}\right)},
%%\frac{\cosh{\left(L\sqrt{2\eta/D}\right)}}{\sinh{\left(L\sqrt{2\eta/D}\right)}},
\ee
where we used the well-known identity
\[
\sum_{n=1}^{\infty}\frac{1}{n^2+b^2}
=
\frac{\pi}{2b}
\coth{(\pi b)}-\frac{1}{2b^2} .
%%\frac{\cosh{(\pi b)}}{\sinh{(\pi b)}}-\frac{1}{2b^2} .
\]
Finally, upon using the notations $g_L=\xi/L$, with $\xi =2\pi \nu D$ and $E_\text{Th}=D/L^2$, the propagator (\ref{propag_1d}) implies the expression $c(\eta)=\frac{1}{2}g_L f(2\eta/E_\text{Th})$ with $f(z)=\sqrt{z}\tanh\sqrt{z}$ mentioned in the main text.
It also naturally coincides with the function $C(t)$ obtained in (\ref{alpha-res}) by the transfer-function method. Indeed, these two functions are essentially the same object, with the only difference that
$c(\eta)$ emphasizes the dependence on absorption (imaginary frequency) $\eta$ (eventually replaced with the resonance width $\Gamma$) whereas $C(t)$ emphasizes the dependence on the length of the sample $t$ measured in units of the localization length.

\subsection{$d\ge 2$}

To find the diffusion propagator in cubic sample of any dimension, one starts by representing it into the sum over discrete Fourier modes as
\be\label{diff_prop_d_sum}
\Pi_d(0,0,\eta)\approx  \frac{1}{2\pi\nu\eta L^d}+\frac{1}{\pi\nu L^d}\sum_{q_1,\ldots,q_d}
\frac{1}{D(q_1^2+\ldots q_d^2)+2\eta}
\ee
where we explicitly separated the zero mode contribution, so that in a cube of linear size $L$ the summation goes over positive integers $n_i=1,2,\ldots, n_\text{max}$ which parametrize mode vectors $q_i=\frac{\pi n_i}{L}, i=1,\ldots, d$. The cutoff parameter $n_\text{max}$ should be chosen to ensure $\text{max}(q_i)\sim \pi/l$, with
$l$ being the mean free path due to underlying disordered potential, in order to stay inside the range of validity of the diffusion approximation. The good metal perturbation theory is controlled by the two large parameters: dimensionless conductance $g_L=2\pi \nu DL^{d-2}=2\pi E_{Th}/\Delta\gg 1$ and  the
large effective absorption/resonance widths $y=\frac{2\pi\eta}{\Delta}\gg 1$. We also will consider samples of the linear length $L$ much exceeding the mean free path $l$, which implies $E_c\ll \tau^{-1}$, with $\tau=l^2/D$ being the mean-free time.
Now passing to continuum limit by replacing the summation by integration, and using spherical coordinates in the Fourier space implies
\be\label{diff_prop_d_cont}
\Pi_d(0,0,\eta)\approx  \frac{1}{\pi\nu}\left[\frac{1}{2\eta L^d}+\frac{1}{\pi^{d/2}2^{d-1}\Gamma(d/2)}\int_{|q|=\frac{\pi}{L}}^{|q|=\frac{\pi}{l}}
\frac{|q|^{d-1}\,dq}{D|q|^2+2\eta}\right]
\ee
In particular, for the most interesting cases $d=2$ and $d=3$ one can evaluate the integral explicitly. In particular,
\be\label{diff_prop_d=2}
\Pi_{d=2}(0,0,y)\approx  \frac{1}{y}+\frac{1}{2\pi g_L}\ln\left(\frac{y+\pi a}{y+\pi b}\right),
\quad a=\frac{\pi^2}{\Delta\tau}, \, b=\frac{\pi^2 E_{Th}}{\Delta}=\pi g_L/2
\ee
so that $b/a=E_{Th}\tau\ll 1$. Note that in the parametrically big range  $1 \ll y\ll g_L$ the expression
(\ref{diff_prop_d=2}) takes the form
\be \label{diff_prop_d=2a}
\Pi_{d=2}(0,0,y)\approx  \frac{1}{y}+\frac{1}{C_2}, \quad  C_2=\frac{\pi g_L}{\ln\left(\frac{L}{l}\right)}
\ee

Similarly, for $d=3$ we have:
\be\label{diff_prop_d=3}
\Pi_{d=3}(0,0,y)\approx  \frac{1}{y}+\frac{1}{\pi}\left(\frac{1}{g_l}-\frac{1}{g_L}\right)+\frac{(2y)^{1/2}}{\pi^2g_L^{3/2}}\left[\arctan\left(\frac{1}{\pi}\frac{g_l}
{g_L}\sqrt{\frac{2y}{g_L}}\right)-\arctan\left(\frac{1}{\pi}\sqrt{\frac{2y}{g_L}}\right)\right],
\ee
where $g_L=2\pi\nu\, D\,L\gg g_l=2\pi\nu\,D\,l$. Hence in the range  $1 \ll y\ll g_L$ we have
\be \label{diff_prop_d=3a}
\Pi_{d=3}(0,0,y)\approx  \frac{1}{y}+\frac{1}{C_3}, \quad  C_3=\pi g_l
\ee
Substituting (\ref{diff_prop_d=3a}) into (\ref{den_pert_1c}) yields the resonance density provided by the formula
Eq. (\ref{rho-3D-diff_text}) of the main text.

\end{document}